\documentclass[preprints,article,accept,moreauthors,pdftex]{Definitions/mdpi} 

\usepackage{bm}
\usepackage{bbm}
\usepackage{listings}
\usepackage{subfigure}
\usepackage{amsmath}
\usepackage{amssymb}
\usepackage{booktabs}
\usepackage{hyperref}
\usepackage{fourier}
\usepackage[T1]{fontenc}
\usepackage[]{avant}
\usepackage{longtable}
\usepackage[ruled, lined, linesnumbered, commentsnumbered, longend]{algorithm2e}

\hypersetup{colorlinks,citecolor=blue}

\firstpage{1} 
\pubvolume{xx}
\issuenum{1}
\articlenumber{1}
\pubyear{2023}
\copyrightyear{2023}
\history{}

\newcommand*\diff{\mathop{}\!\kern0pt\mathrm{d}}
\newcommand{\iu}{\mathrm{i}\mkern1mu}

\Title{The Quadratic Local Variance Gamma Model: an arbitrage-free interpolation of class C3 for option prices}
\Author{Fabien Le Floc'h}

\AuthorNames{Fabien Le Floc'h}

\address{fabien@2ipi.com}
\abstract{This paper generalizes the local variance gamma model of Carr and Nadtochiy, to a piecewise quadratic local variance function. The formulation encompasses the piecewise linear Bachelier and piecewise linear Black local variance gamma models. The quadratic local variance function results in an arbitrage-free interpolation of class $\mathcal{C}^3$. The increased smoothness over the piecewise-constant and piecewise-linear representation allows to reduce the number of knots when interpolating raw market quotes, thus providing an interesting alternative to regularization while reducing the computational cost. }
\keyword{volatility surface; interpolation; risk-neutral density; arbitrage; quantitative finance; European options}

\begin{document}
	
	\section{Introduction}
The financial markets provide option prices for a discrete set of strike prices and maturity dates. In order to price over-the-counter vanilla options with different strikes, or to hedge complex derivatives with vanilla options, it is useful to have a continuous arbitrage-free representation of the option prices, or equivalently of their implied volatilities.  For example, the variance swap replication of Carr and Madan consists in integrating a specific function over a continuum of vanilla put and call option prices \citep{carr2001towards,carr2008robust}. More generally, \citet{breeden1978prices} have shown that any path-independent claim can be valued by integrating over the probability density implied by market option prices. An arbitrage-free representation is also particularly important for the Dupire local volatility model \citep{dupire1994pricing}, where arbitrage will translate to a negative local variance. A option price representation of class $\mathcal{C}^2$ is also key to guarantee the second-order convergence of numerical schemes applied to the Dupire partial differential equation, commonly used to price exotic financial derivative contracts. 

A rudimentary, but popular representation is to interpolate market implied volatilities with a cubic spline across option strikes. Unfortunately this may not be arbitrage-free as it does not preserve the convexity of option prices in general. A typical convex interpolation of the call option prices by quadratic splines or rational splines is also not satisfactory in general since it may generate unrealistic oscillations in the corresponding implied volatilities, as evidenced in \citep{jackel2014clamping}.  \citet{kahale2004arbitrage} designs an arbitrage-free interpolation of the call option prices, which however requires convex input quotes, employs two embedded non-linear minimizations, and it is not proven that the algorithm for the interpolation function of class $\mathcal{C}^2$ converges. 


More recently, \citet{andreasen2011volatility} have proposed to calibrate the discrete piecewise constant local volatility corresponding to a single-step finite difference discretization of the forward Dupire equation. In their representation of the local volatility, the authors use as many constants as the number of market option strikes for an optimal fit. It is thus sometimes considered to be "non-parametric". Their technique works well in general but requires some care around the choice of discretization grid: it must be sufficiently dense so that two market strikes do not fall in between the same consecutive grid nodes, and sufficiently wide to properly model the boundary behaviour. Those two requirements complicate, and slow down the non-linear optimization involved in the technique. Furthermore the output is a discrete set of option prices, which, while relatively dense, must still be interpolated carefully to obtain the price of options whose strike falls in between grid nodes.

\citet{lefloch2019model2} derived a specific B-spline collocation to fit the market option prices, while ensuring the arbitrage-free property at the same time. While the fit is quite good in general, it may not be applicable to interpolate the original quotes with high accuracy. For example, input quotes may already be smoothed out if they stem from a prior model, or from a market data broker, or from another system in the bank. In those cases, it is desirable to use a nearly exact interpolation. 

\citet{lefloc2020arbitrage} extends the local variance gamma model of \citet{carr2017local}, which relies on a piecewise-constant representation of the local variance function, to a piecewise-linear Bachelier representation. This paper generalizes the model to a piecewise-quadratic function. It encompasses the piecewise-linear Bachelier and piecewise-linear Black representations. The full piecewise-quadratic model results in an arbitrage-free interpolation of class $\mathcal{C}^3$ for the option prices. The smoother implied probability density allows for the use of a sparser set of interpolation knots, thus providing an alternative to  regularization in order to avoid overfitting. In addition, a sparser set of knots reduces the computational cost of the technique.

\section{Dupire's PDDE in the local variance gamma model}
We recall Dupire's partial difference differential equation (PDDE) for a call option price $C(T,x)$ of strike $x$ and maturity $T$ \citep{carr2017local}:
\begin{equation}
	\frac{C(T,x)-\max(X(0)-x,0)}{T} =  \frac{1}{2}a^2(x)  \frac{\partial^2 C(T,x) }{\partial x^2} \,,\label{eqn:dupire_pdde}
\end{equation}
for a Martingale asset price process $X(t)$ of expectation $\mathbb{E_\mathbb{Q}}[X(t)] = X(0)$. $X(t)$ is driftless, and may thus be the the time $T$-forward value of a stock price, a foreign exchange rate or a commodity price at time $t$. 

Let $\{ x_0, x_1,...,x_m, x_{m+1} \}$  be a increasing set of the strike prices, such that $x_0=L$, $x_{m+1}=U$ with the interval $(L,U)$ being the spatial interval where the  asset $X$ lives. Furthermore, we require the following to hold
\begin{equation*}\exists s \in \lbrack 1,m \rbrack | x_s = X(0)\,.\end{equation*}   
The $(x_1,...,x_m)$ may correspond to the strike prices of the options of maturity $T$ we want to calibrate against, along with the forward price as in \citep{carr2017local, lefloc2020arbitrage}. This choice allows for a nearly exact fit. It may also be some specific discretization of size $m$ with $m$ lower or equal to the number of market strike prices.

We consider $a$ to be a piecewise-quadratic function of class $\mathcal{C}^0$ on $[x_0,x_{m}]$.

Let $V$ be the function defined by $V(x) = C(T,x)- \max(X(0)-x,0)$. $V$ is effectively the time-value\footnote{We consider undiscounted option prices.} of the option. That is, if the option is out-of-the-money (OTM), then $V(x)$ is the option value. If the option  is in-the-money (ITM), then $V(x)$ is the option value minus the intrinsic value. The Dupire PDDE leads to
\begin{equation}
	V(x) =  \frac{1}{2}a^2(x) T  \left[V''(x)+\delta\left(x-X(0)\right)\right] \,,\label{eqn:otm_ode_dirac}
\end{equation}
on the interval $(L,U)$, where $\delta$ is the Dirac delta function. Instead of solving Equation \ref{eqn:otm_ode_dirac} directly, we look for a solution $V$ on the two intervals  $(L, X(0))$ and $(X(0),U)$ separately. On each interval, we have
\begin{equation}
	V(x) =  \frac{1}{2}a^2(x) T  V''(x) \,,\label{eqn:otm_ode}
\end{equation}
Then, the continuity of $\frac{\partial C}{\partial x}$ at $x=X(0)$ implies 
\begin{equation}\lim\limits_{x \to X(0)-} V'(x) = 1 + \lim\limits_{x \to X(0)^+} V'(x)\,.\label{eqn:jump_cond_v}\end{equation}
In order to define a unique $V$, we also impose the absorbing boundary conditions
\begin{equation}
	V(L) = 0 = V(U)\,.\label{eqn:boundary_v}
\end{equation}
The assumption is that the option value is zero for OTM and the
intrinsic value for ITM. For these conditions to be plausible, it must be that
$L$ and $U$ are far below and above, respectively, the maturity forward price. 

The continuity of the second derivative of the call option price $C(T,x)$ at $x=X(0)$ follows from the continuity of $a(x)$  everywhere, including $x=X(0)$ as per \citet[Theorem 3.3]{carr2017local}. We may further impose a $\mathcal{C}^1$ continuity of the probability density of $X$ on $(L,U)$, or equivalently a  $\mathcal{C}^3$  continuity of the option prices $C$. This requires $a(x)$ to be  $\mathcal{C}^1$ on $(L,X(0))$ and $(X(0),U)$ along with the following continuity condition at $x=x_s$:
\begin{equation}
	\lim\limits_{x \to X(0)-}	\left(\frac{V}{a^2}\right)' (x) = \lim\limits_{x \to X(0)+}\left(\frac{V}{a^2}\right)' (x) \,.\label{eqn:lvg_c3_condition}
\end{equation}
Note that $a'(x)$ is not necessarily continuous, and actually should not be continuous for $C$ to be $\mathcal{C}^3$. In particular when $a(x)=1$, $C$ is not $\mathcal{C}^3$, because we can not impose Equation \ref{eqn:lvg_c3_condition}, given that it would contradict Equation \ref{eqn:jump_cond_v}.


 

\section{Explicit solution}\label{sec:pdde_solution}
Let $a(x) = \alpha_i x^2 + \beta_i x + \gamma_i$ on $[x_i, x_{i+1})$ with $(\alpha_i, \beta_i, \gamma_i) \in \mathbb{R}^3$. Being a quadratic, $a$ may also be expressed as $a(x) = \alpha_i (x-\tilde{x}_{i,1})(x-\tilde{x}_{i,2})$ with 
\begin{align*}
	\tilde{x}_{i,1} =  \frac{-\beta_i + \sqrt{\delta_i}}{2\alpha_i}\,,& \quad
		\tilde{x}_{i,2} =  \frac{-\beta_i - \sqrt{\delta_i}}{2\alpha_i}\,,\quad \textmd{ with } \delta_i = \beta_i^2 - 4\alpha_i \gamma_i\,,\quad \textmd{ for } \alpha \neq 0\,.
\end{align*}
In particular, $\tilde{x}_{i,1}$ and $\tilde{x}_{i,2}$ may be complex numbers.
When $\alpha_i = 0$ and $\beta_i \neq 0$, we may define $\delta_i = \beta_i^2$ and we have  $a(x) = \beta_i (x-\tilde{x}_{i,1})$ with $\tilde{x}_{i,1} = - \gamma_i / \beta_i$.

The solutions of Equation \ref{eqn:otm_ode} on $[x_i,x_{i+1})$ read
\begin{align}
	V(x) &= \frac{\chi_i(x)}{\chi_i(x_i)} \left[\Theta_i^c \cosh\left(\omega_i\left(z_i(x)-z_i(x_i)\right)\right) + \Theta_i^s \sinh\left(\omega_i\left(z_i(x)-z_i(x_i)\right)\right)\right]\,,\label{eqn:lvg_V} 
\end{align}
with\footnote{See Appendix \ref{sec:avoid_complex_numbers} on how to avoid the use of complex numbers.}
\begin{align*}z_i(x)&= \ln \left(\frac{x - \tilde{x}_{i,1}}{x - \tilde{x}_{i,2}}\right)\,,&\quad \omega_i &=\frac{1}{2} \sqrt{1+ \frac{8}{\delta_i T}} \,,&\quad \chi_i &=\sqrt{\left(x - \tilde{x}_{i,1}\right)\left(x - \tilde{x}_{i,2}\right)}\,,&\quad
\textmd{ for }\alpha_i \neq 0\,,\\
z_i(x)&= \ln \left|x - \tilde{x}_{i,1}\right|\,,&\quad  \omega_i &=\frac{1}{2} \sqrt{1+ \frac{8}{\delta_i T}} \,,&\quad \chi_i &=\sqrt{\left|x - \tilde{x}_{i,1}\right|}\,,&\quad \textmd{for } \alpha_i = 0, \textmd{ and } \beta_i \neq 0\,,\\
	z_i(x) &= x\,,&\quad \omega_i &= \frac{1}{\gamma_i}\sqrt{\frac{2}{T}}\,,&\quad \chi_i &=1\,,&\quad \textmd{for } \alpha_i = 0, \textmd{ and } \beta_i = 0\,.	
\end{align*}
where $(\Theta_i^c, \Theta_i^s) \in \mathbb{C}^2$. The normalization makes $V(x_i) = \Theta_i^c$.

The derivative of $V$ reads
\begin{align}
	V'(x) &= \frac{\chi_i(x)}{\chi_i(x_i)} z_i'(x) \left[(\kappa_i\Theta_i^c + \omega_i\Theta_i^s) \cosh\left(\omega_i(z_i(x)-z_i(x_i))\right) + (\kappa_i\Theta_i^s + \omega_i\Theta_i^c) \sinh\left(\omega_i(z_i(x)-z_i(x_i))\right)\right]\,, 
\end{align}
with
\begin{align*}
	z_i'(x)	 &=	\frac{1}{x-\tilde{x}_{i,1}} - \frac{1}{x-\tilde{x}_{i,2}}\,,&\quad \kappa_i &= \frac{1}{2z_i'(x)}\left(\frac{1}{x-\tilde{x}_{i,1}} + \frac{1}{x-\tilde{x}_{i,2}}\right)\,,&\quad \textmd{ for }  \alpha_i \neq 0\,,\\
	z_i'(x)	 &=	\frac{1}{x-\tilde{x}_{i,1}}\,,&\quad \kappa_i &= \frac{1}{2}\,,&\quad \textmd{ for }  \alpha_i = 0 \textmd{ and } \beta_i \neq 0\,,\\
	z_i'(x)	 &=	1\,,&\quad \kappa_i &= 0\,,&\quad \textmd{ for }  \alpha_i = 0 \textmd{ and } \beta_i = 0\,.
\end{align*}


The conditions to impose continuity of $V$ and its derivative at $x=x_{i+1}$ results in the following linear system
\begin{align}
	\cosh_i \Theta_{i}^c +  \sinh_i \Theta_{i}^s &= \frac{\Theta_{i+1}^c}{\chi_i(x_{i+1})}\,, \label{eqn:price_cont}\\
	(\kappa_i \cosh_i + \omega_i \sinh_i)  \Theta_i^c  + (\omega_i  \cosh_i + \kappa_i \sinh_i)\Theta_i^s &= 
	\frac{\left(\kappa_{i+1}\Theta_{i+1}^c+ \omega_{i+1}\Theta_{i+1}^s\right)z_{i+1}'(x_{i+1})}{\chi_i(x_{i+1})z_i'(x_{i+1})} \label{eqn:der_cont}
\end{align}
for $i=0,...,s-2$,
with \begin{align*}
	\cosh_i = \cosh\left(\omega_i(z_i(x_{i+1})-z_i(x_i))\right)\,,&\quad \sinh_i = \sinh\left(\omega_i(z_i(x_{i+1})-z_i(x_i))\right)\,.
\end{align*}

The boundary condition at $x=x_0=L$ translates to $\Theta_0^c = 0$. At $x=x_{m+1}=U$, the boundary condition translates to $\Theta_{m}^c = -\Theta_{m}^s \frac{\sinh_m}{\cosh_m}$.
The jump condition at $x=s$ reads
\begin{align*}
	V_{s-1}(x_s) &= V_s(x_s)\,,\\
	V_{s-1}'(x_s) &= 1+ V_s'(x_s)\,,
\end{align*}
with 
\begin{align*}
	V_{s-1}(x_s) &= \chi_{s-1}(x_s) (\Theta_{s-1}^c \cosh_{s-1} + \Theta_{s-1}^s \sinh_{s-1}) \,,\\
	V_s(x_s) &= \Theta_s^c\,,\\
	V_{s-1}'(x_s) &=\chi_{s-1}(x_s) z_{s-1}'(x_s) \left[ (\kappa_{s-1} \Theta_{s-1}^c + \omega_{s-1} \Theta_{s-1}^s)\cosh_{s-1} +  (\omega_{s-1} \Theta_{s-1}^c + \kappa_{s-1} \Theta_{s-1}^s) \sinh_{s-1}\right] \,,\\
	V_s'(x_s) &= (\kappa_s\Theta_s^c + \omega_s\Theta_s^s) z_s'(x_s)\,.
\end{align*}

From the above equations, we deduce that the coefficients $\Theta_i^c,\Theta_i^s$ are solutions of the following tridiagonal system\footnote{the $A,B$ and $C$ of the tridiagonal matrix are not known a-priori. In the case of a quadratic B-spline representation, they will be a function of the B-spline parameters $\bm\lambda$. 
    The value $\bm\lambda$ will be determined by a least squares minimization of 
    the error between model prices and market prices. One step in the least-squares objective function is to solve the (then known) tridiagonal system.} 
\begin{equation}\label{eqn:tridiagonal_lvg}
	\begin{pmatrix}
		B_0 & C_0 &  & 0 \\
		A_1 & \ddots & \ddots  \\
		& \ddots & \ddots & C_{2m} \\
		0 &  & A_{2m+1} & B_{2m+1} \end{pmatrix}  \begin{pmatrix} \Theta_0^s \\
		\Theta_0^c\\
		\vdots \\
		\Theta_m^s\\
		\Theta_m^c \end{pmatrix}   =   \begin{pmatrix} D_0 \\
		\vdots \\
		D_{2m+1}  \end{pmatrix} \,,
\end{equation}
with $D_i = 0$ for $i \notin \left\{2s-1,2s\right\}$, $D_{2s-1} = D_{2s} = 1$, 
\begin{align*}\begin{cases}
	A_{2i+1} &= (\omega_i  \cosh_i + \kappa_i \sinh_i) \chi_i  z_i'(x_{i+1}) - \kappa_{i+1} \sinh_i \chi_i z_{i+1}'(x_{i+1})\,,\\ 
	B_{2i+1} &= (\kappa_i  \cosh_i + \omega_i  \sinh_i)  \chi_i  z_i'(x_{i+1}) - \kappa_{i+1}  \cosh_i  \chi_i  z_{i+1}'(x_{i+1}) \,,\\
	C_{2i+1} &=  -\omega_{i+1}  z_{i+1}'(x_{i+1})\,,\\ 
A_{2i+2} &= \left(\kappa_i  \cosh_i + \omega_i  \sinh_i - \frac{\omega_i \cosh_i + \kappa_i \sinh_i}{\sinh_i} \cosh_i\right)  \chi_i  z_i'(x_{i+1}) \,,\\
B_{2i+2} &=  -\omega_{i+1}  z_{i+1}'(x_{i+1})\,,\\ 
	C_{2i+2} &= \frac{\omega_i  \cosh_i + \kappa_i \sinh_i}{\sinh_i}  z_i'(x_{i+1}) - \kappa_{i+2}  z_{i+1}'(x_{i+1})\,,\end{cases}
\end{align*}
for $i=0,...,m-1$, and $B_0 = 0, C_0=1, A_{2m+1}=\sinh_{m},   B_{2m+1} = \cosh_{m}$.


Using the continuity of $V(x_s)=\Theta^c_s$, the jump condition of $V'$ at $x_s$ and the continuity of $a(x_s)$, the  $\mathcal{C}^3$ condition (Equation \ref{eqn:lvg_c3_condition}) reads
\begin{align}  1 - 2 \Theta^c_s\frac{\lim\limits_{x \to x_s-}a'(x)}{a(x_s)}    =  - 2\Theta^c_s \frac{\lim\limits_{x \to x_s+} a'(x)}{a(x_s)}\,.\label{eqn:lvg_c3_condition_a}\end{align}
Equation \ref{eqn:lvg_c3_condition_a} implies that $a'$ is not continuous at $x_s$, unless $a(x_s)=0$. The condition can not be imposed as an additional constraint on $\Theta_s^c$ since its value is already fully determined by the tridiagonal system. It may however be imposed by choosing the correct model parameter to adjust the value of $a$ at $x_s$ along with its left and right derivative values.

\section{Parameterizations}
\subsection{Linear Bachelier}
The linear Bachelier local variance consists in $\alpha_i=0$ and may be rewritten using  values at the knots $\sigma_i$ as
\begin{align}
	a(x) &= \frac{x-x_i}{x_{i+1}-x_i} (\sigma_{i+1}-\sigma_i) + \sigma_i \quad \textmd{ for } x_i \leq x < x_{i+1}\,,\quad i=0,...,m\,,
\end{align}
where the parameters $\sigma_i > 0$.

It corresponds to the parameterization studied in \citep{lefloc2020arbitrage}, where it is shown that the local variance function must not be $\mathcal{C}^1$ at $x=x_s$ but must follow $\mathcal{C}^3$ condition (Equation \ref{eqn:lvg_c3_condition_a}) in order to avoid a spurious spike at $x=x_s$. Under the linear Bachelier local variance, the condition reads
\begin{equation*}
	 1- 2\Theta_s^c \frac{\sigma_s - \sigma_{s-1}}{(x_s-x_{s-1})\sigma_s}  = - 2\Theta_s^c \frac{\sigma_{s+1} - \sigma_{s}}{(x_{s+1}-x_{s})\sigma_s} \,,
\end{equation*}
or equivalently
\begin{equation}
	\sigma_s = \frac{2\Theta_s^c \left( \frac{\sigma_{s-1}}{x_s - x_{s-1}} + \frac{\sigma_{s+1}}{x_{s+1}-x_s} \right)}{2\Theta_s^c \left( \frac{1}{x_s - x_{s-1}} + \frac{1}{x_{s+1}-x_s} \right) - 1}\,.\label{eqn:alpha_s_iter_bachelier}
\end{equation}
This is not a linear problem, as $\Theta_s^c$ depends on $\sigma_s$ through $\Theta_{s+1}^c, \Theta_{s+1}^s$ in a non-linear way (Equations \ref{eqn:price_cont} and \ref{eqn:der_cont}). Starting with the algorithm described in Section \ref{sec:pdde_solution} to compute $\Theta^c, \Theta^s$, using Equation \ref{eqn:alpha_s_iter_bachelier} with $\Theta_s^c \approx V_{\textsf{market}}(x_s)$ as initial guess for $\sigma_s$, we may however apply the following iteration 
\begin{itemize}
	\item Update $\sigma_s$ through Equation \ref{eqn:alpha_s_iter_bachelier}.
	\item Recalculate $\Theta_i^c$ and $\Theta_i^s$ for $i=0,...,m$ by solving the updated tridiagonal system.
\end{itemize}
 Three iterations are enough in practice. It may happen that the denominator of the right hand side of
 Equation \ref{eqn:alpha_s_iter_bachelier} is negative when the distances $x_{s+1}-x_s$ and $x_s - x_{s-1}$ are large. We found it sufficient to move $x_{s+1}$ and $x_{s-1}$ such that the denominator is guaranteed\footnote{We pick $h= \min\left(x_{s+1}-x_s,x_{s}-x_{s-1},3 V_{\textsf{market}}(x_s)\right)$ and set the new $x_{s+1}=x_s+h$ $x_{s-1}=x_s-h$.} to be positive for the initial guess.

\subsection{Linear Black}

The linear Black local variance model is defined by $\gamma=0$. The local variance function may be rewritten using  values at the knots $\sigma_i$ as
\begin{align}
	a(x) &= \left(\frac{x-x_i}{x_{i+1}-x_i} (\sigma_{i+1}-\sigma_i) + \sigma_i\right)x \quad \textmd{ for } x_i \leq x < x_{i+1}\,,\quad i=0,...,m\,,
\end{align}
where the parameters $\sigma_i > 0$.
Interestingly, the $\mathcal{C}^3$ condition (Equation \ref{eqn:lvg_c3_condition_a}) is also given by Equation \ref{eqn:alpha_s_iter_bachelier}.

\subsection{Positive quadratic B-spline}
A B-spline parameterization with positive coefficients implies $a$ positive. Furthermore, Equation \ref{eqn:lvg_c3_condition_a} imposes a double knot at $x=x_s$ (because the derivative of $a$ is not continuous there). We thus consider
\begin{equation}
	a(x) = \sum_{i=1}^{m+3} \lambda_i B_{i,3}(x)
\end{equation}
where $\lambda_i > 0$ and $B_{i,3}$ is the quadratic basis spline with knots $\bm{t} =  (L,L,L,x_1,x_2,...,X(0),X(0),...,x_{m},U,U,U)$. In particular, we have $t_{s+2}=t_{s+3}=X(0)$. Using the B-spline derivative identity \citep{de1978practical} and the fact that the order of the B-spline is 3, we obtain
\begin{equation*}
	a'(x) = 2 \sum_{i} \frac{\lambda_i - \lambda_{i-1}}{t_{i+2}-t_{i}}B_{i-1,2}(x)\,,
\end{equation*}
and the $\mathcal{C}^3$ condition reads
\begin{align*}
	\sum_i \lambda_i B_{i,3}(x_s) = 4\Theta^c_s \left(\frac{\lambda_{s+1}-\lambda_{s}}{t_{s+3}-t_{s+1}} B_{s,2}(x_s^-)- \frac{\lambda_{s+2}-\lambda_{s+1}}{t_{s+4}-t_{s+2}}B_{s+1,2}(x_s^+)\right)\,.
\end{align*}
Using the definitions of $B_{i,3}$ and $B_{i,2}$ we obtain
\begin{equation*}
	\lambda_{s+1} = 4\Theta^c_s \left(\frac{\lambda_{s+1}-\lambda_{s}}{t_{s+3}-t_{s+1}} - \frac{\lambda_{s+2}-\lambda_{s+1}}{t_{s+4}-t_{s+2}}\right)\,,
\end{equation*}
or equivalently
\begin{equation}
	\lambda_{s+1} =  \frac{4\Theta^c_s \left(\frac{\lambda_{s}}{t_{s+3}-t_{s+1}} + \frac{\lambda_{s+2}}{t_{s+4}-t_{s+2}}\right)}{4\Theta^c_s\left(\frac{1}{t_{s+3}-t_{s+1}} + \frac{1}{t_{s+4}-t_{s+2}}\right) -1}\,.
\end{equation}
\begin{figure}[H]
	\centering{
		\includegraphics[width=.75\textwidth]{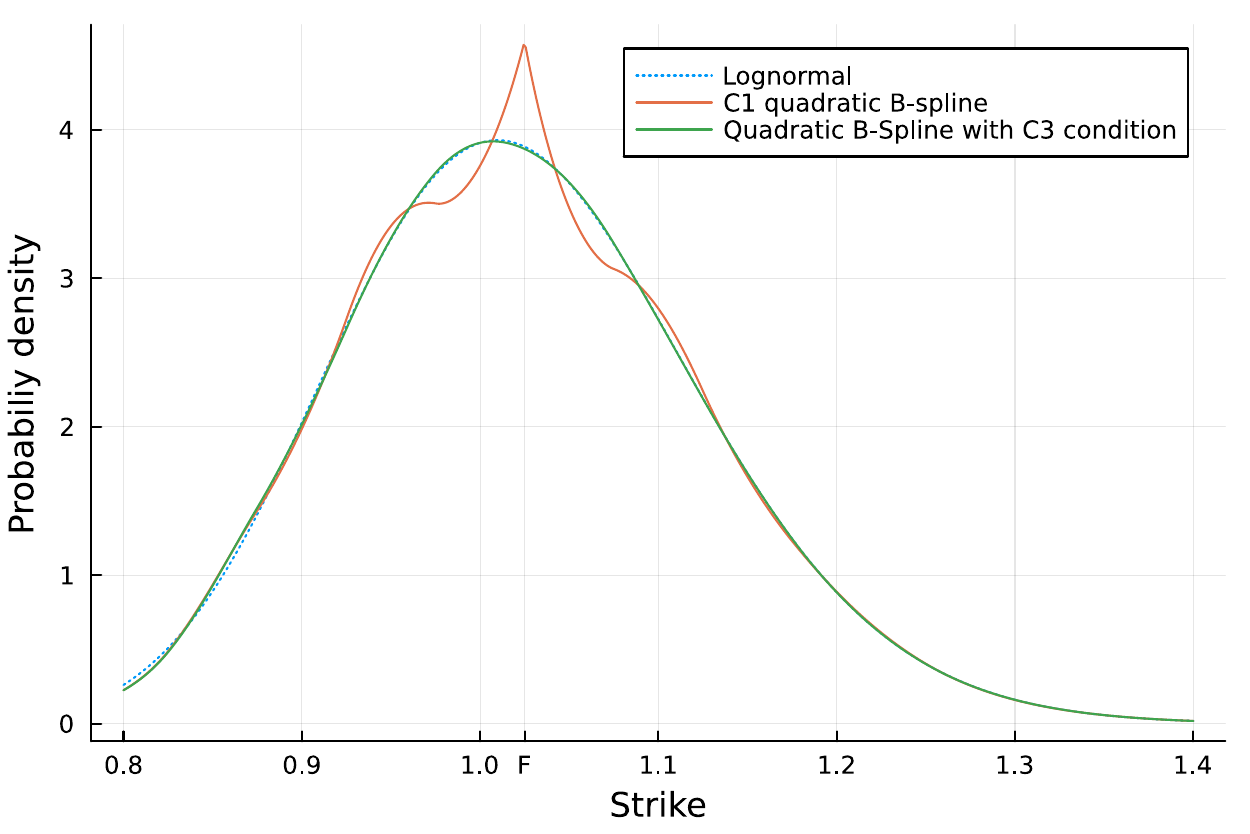}}
	\caption{Implied probability density for the quadratic LVG model with or without the $\mathcal{C}^3$ condition (Equation \ref{eqn:lvg_c3_condition}), fitted to a Black-Scholes model with constant volatility $\sigma_B=20\%$, time to expiry $T=0.25$ and forward $1.025$.\label{fig:lvgq_lognormal_density}}
\end{figure} 
As an illustration, we consider the same example as in \citep{lefloc2020arbitrage}: we fit the quadratic local variance gamma (LVG) model to 10 option prices of strikes (0.85, 0.90, 0.95, 1, 1.05, 1.1, 1.15, 1.2, 1.3, 1.4), obtained by the Black-Scholes model with constant volatility $\sigma_B=20\%$, time to maturity $T=0.25$ and forward price $1.025$. We know that the theoretical distribution is a lognormal distribution. A straightforward $\mathcal{C}^1$ quadratic B-spline leads to a large spike in the probability density implied from the calibrated LVG model (Figure \ref{fig:lvgq_lognormal_density}). Adding the $\mathcal{C}^3$ condition through an additional B-spline knot recovers a smooth implied probability density.

\section{Calibration}
\subsection{Error measure}
The calibration of a single maturity consists in finding the parameters (the $\bm{\alpha}$ for the linear models, or $\bm{\lambda}$ for the quadratic B-spline) such that the function $C(T,x)$, solution of the Dupire PDDE fits the market option prices $(\hat{C}_i)_{i=1,...n}$ of respective strikes $(K_i)_{i=1,...,n}$ according to an appropriate measure.
A common practice is to perform a least-squares minimization of the error measure $E$ defined by
\begin{equation}\label{eqn:objective}
	E = \sum_{i=1}^m \mu_i^2\left(\sigma(\alpha, x_i) - \hat{\sigma}_i\right)^2\,,
\end{equation}
with $\alpha_i > 0$ for $i=1,...,m$ and where $\sigma(\alpha, x)$ is the implied volatility corresponding to the option prices obtained with the piecewise-linear local gamma variance model and $\hat{\sigma}_i$ is the market implied volatility at strike $x_i$, $(\mu_i)_{i=1,...,m}$ are weights associated to the accuracy of the fit at each point. 

In order to solve this non-linear least-squares problem, we will use the Levenberg-Marquardt algorithm as implemented by \citet{klare2013gn}. The box constraints $\alpha_i > 0$ can be added in a relatively straightforward manner to any Levenberg-Marquardt algorithm, through the projection technique described in \citep{kanzow2004levenberg}, or through a variable transform from $\mathbb{R}$ to a subset of $\mathbb{R^+}$ (for example through the function $x \to x^2 + \epsilon$ with some small positive $\epsilon$). 

The implied volatility for a given option price may be found efficiently and accurately through the algorithm of \citet{jackel2013let}. In general, we prefer to solve an almost equivalent formulation in terms of option prices, using the error measure $E_V$ defined by
\begin{equation}\label{eqn:objective_price}
	E_V = \sum_{i=1}^m w_i^2\left(C(\alpha, x_i) - \hat{C}_i\right)^2\,,
\end{equation}
with $C(\alpha, x)$ being the local variance gamma option price with parameter $\alpha$ and strike $x$, and the capped inverse Vega weights $w_i$ given by
\begin{equation}
	w_i = \min\left(\frac{1}{\nu_i}, \frac{10^6}{X(0)} \right)\mu_i\,,
\end{equation}
where $\nu_i= \frac{\partial \hat{C}_i}{\partial \sigma}$ is the Black-Scholes Vega corresponding the market option price $\hat{C}_i$, and $10^6$ is a cap applied to avoid numerical issues related to the limited machine accuracy (see \citet{lefloch2019model2,lefloc2020arbitrage} for the justification).

\subsection{Exact interpolation}
Sometimes, it is desirable to interpolate a given set of reference prices nearly exactly. This is typically the case when the reference prices come from some prior model. We apply the same least-square minimization but choose the number of free parameters to be equal to the number of reference prices.

\subsubsection{Linear models}
For the linear models, this means to use $m=n$ and set $\alpha_0 = \alpha_1$ and $\alpha_{m+1}=\alpha_{m}$ to model a flat extrapolation.  
In general, the market strikes will not include $X(0)$. In this case, $X(0)$ must be added to the knots $\{x_i\}_{i=1,...,m}$ used in the local variance gamma representation. This adds one more parameter $\alpha_s$ to the representation, where $s$ is the index corresponding to $X(0)$ in the set of knots. The value of $\alpha_s$ is not free, it is given by the $\mathcal{C}^3$ condition (Equation \ref{eqn:lvg_c3_condition}) and enforced through the iterative procedure described in the previous sections.

\subsubsection{B-spline knots locations}
For the linear Bachelier and Black parameterization, choosing the knots at the market strikes works well. The situation is more complex for the quadratic B-spline parameterization. 

Let $(K_i)_{i=1,...,n}$ be the market options strikes. Let the index $i_F$ be such that $K_{i_F} \leq F < K_{i_F+1}$.
We may:
\begin{itemize}
	\item place the knots at the market strikes (labeled "Strikes" in the figures)
	\begin{align*}
		\bm{t} = \begin{cases}\left(L,L,L, K_{1}, ..., K_{i_F}, F, F, K_{i_F+1},...,K_{n},U,U,U\right) &\textmd{ if }K_{i_F} \neq F\\
			\left(L,L,L, K_{1}, ..., K_{i_F}, F, K_{i_F+1},...,K_{n},U,U,U\right) &\textmd{ if } K_{i_F} = F
		\end{cases}\,,
	\end{align*}  
The dimension of $\bm{\lambda}$ is then $n_\lambda=n + 5$ if $F \neq K_{i_F}$ and $n + 4$ if $F = K_{i_F}$. The change in the number of dimensions suggests that the interpolation may change  significantly when the forward price moves across a market strike.
	\item place the knots in the middle of market strikes. According to \citep[p. 61]{de1978practical}, the $\mathcal{C}^1$ quadratic spline is then solution to a diagonally dominant tridiagonal system, which increases the stability and reduce oscillations of the interpolation. There are however several ways to do it:
	\begin{itemize}
		\item choose the direct mid-points (labeled "Mid-Strikes") \begin{align*}\bm{t} = \begin{cases}\left(L,L,L, \frac{K_{1} + K_{2}}{2}, ..., \frac{K_{i_F-1}+K_{i_F}}{2}, F, F, \frac{K_{i_F}+K_{i_F+1}}{2},...,\frac{K_{n-1}+K_{n}}{2},U,U,U\right) &\textmd{ if } F <\frac{K_{i_F}+K_{i_F+1}}{2} \\
				\left(L,L,L, \frac{K_{1} + K_{2}}{2}, ..., \frac{K_{i_F-1}+K_{i_F}}{2}, \frac{K_{i_F}+K_{i_F+1}}{2},F, F, ...,\frac{K_{n-1}+K_{n}}{2},U,U,U\right)
			&\textmd{ if } F > \frac{K_{i_F}+K_{i_F+1}}{2}\\
			\left(L,L,L, \frac{K_{1} + K_{2}}{2}, ..., \frac{K_{i_F-1}+K_{i_F}}{2}, F, F, \frac{K_{i_F+1}+K_{i_F+2}}{2},...,\frac{K_{n-1}+K_{n}}{2},U,U,U\right)
		&\textmd{ if }  F = \frac{K_{i_F}+K_{i_F+1}}{2}	\end{cases}\,,\end{align*}
The dimension of $\bm{\lambda}$ is then $n_\lambda=n + 4$ if $F \neq K_{i_F}$ and $n + 3$ if $F = K_{i_F}$. 
		\item choose the mid-points, excluding the point closest to the forward price $F$ (labeled "Mid-X") \begin{equation*}\bm{t} = \left(L,L,L, \frac{K_{1} + K_{2}}{2}, ..., \frac{K_{i_F-1}+K_{i_F}}{2}, F, F, \frac{K_{i_F+1}+K_{i_F+2}}{2},...,\frac{K_{n-1}+K_{n}}{2},U,U,U\right)\,,\end{equation*}		
The dimension of $\bm{\lambda}$ is then $n_\lambda=n + 3$.
		\item choose  the mid-points, excluding the point closest to the forward price and placing the first and last strike in the middle of two knots (labeled "Mid-XX")
		\begin{equation*}\bm{t} = \left(L,L,L, \frac{3K_{1} - K_{2}}{2},\frac{K_{1} + K_{2}}{2}, ..., \frac{K_{i_F-1}+K_{i_F}}{2}, F, F, \frac{K_{i_F+1}+K_{i_F+2}}{2},...,\frac{K_{n-1}+K_{n}}{2},\frac{3K_{n}-K_{n-1}}{2},U,U,U\right)\,,\end{equation*}
The dimension of $\bm{\lambda}$ is then $n_\lambda=n + 5$.
	\end{itemize}
 \item use a uniform discretization of $[K_1,K_n]$ composed of $n+1$ points, and shift it such that the forward is exactly part of the knots and we have $n_\lambda=n+5$.
\end{itemize}
In each of those case, we make sure to add the forward price as a double knot, as well as the boundaries $L,U$. The dimension of $\bm{\lambda}$ implied by the knots is larger than the number of market strikes. We choose the extra parameters as such:
\begin{itemize}
	\item if $n_\lambda = n+5$, we set $\lambda_1 = \lambda_2 = \lambda_3$, $\lambda_{n+3}=\lambda_{n+4}=\lambda_{n+5}$ and $\lambda_{i_F+3}$ is obtained from $\lambda_{i_F+2}$ and $\lambda_{i_F+4}$.
	\item if $n_\lambda = n+4$, we set $\lambda_1 = \lambda_2 = \lambda_3$, $\lambda_{n+3}=\lambda_{n+4}$ and $\lambda_{i_F+3}$ is obtained from $\lambda_{i_F+2}$ and $\lambda_{i_F+4}$.
	\item if $n_\lambda = n+3$, we set $\lambda_1 = \lambda_2$, $\lambda_{n+2}=\lambda_{n+3}$ and $\lambda_{i_F+2}$ is obtained from $\lambda_{i_F+1}$ and $\lambda_{i_F+3}$.
\end{itemize} 

\begin{figure}[H]
		\subfigure[\label{fig:lvgq_lognormal_density_set_A}Set A.]{
			\includegraphics[width=0.5\textwidth]{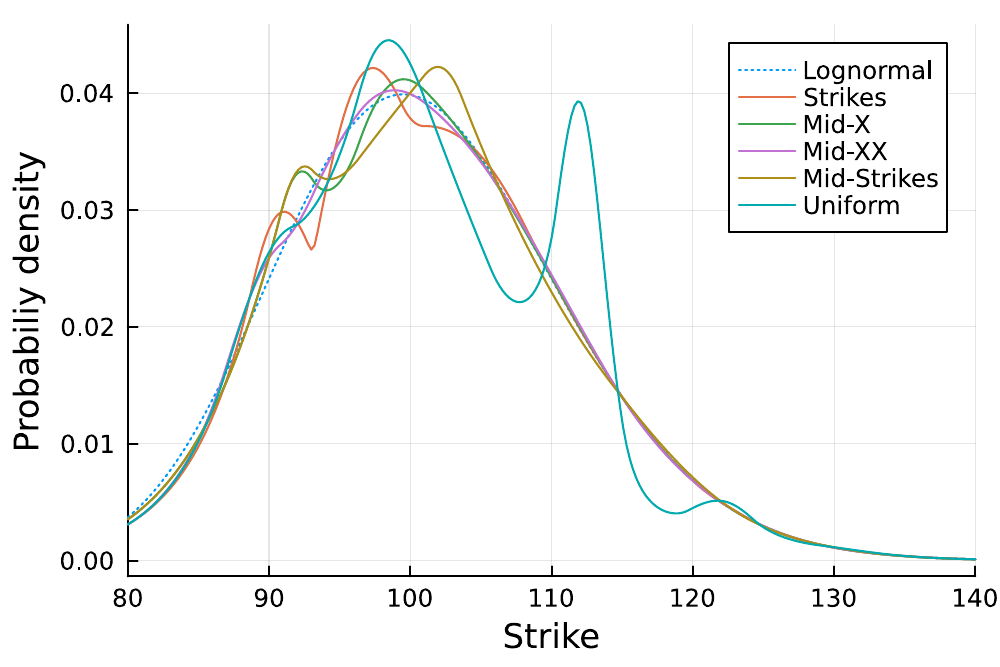}}
		\subfigure[\label{fig:lvgq_lognormal_density_set_B}Set B.]{
			\includegraphics[width=0.5\textwidth]{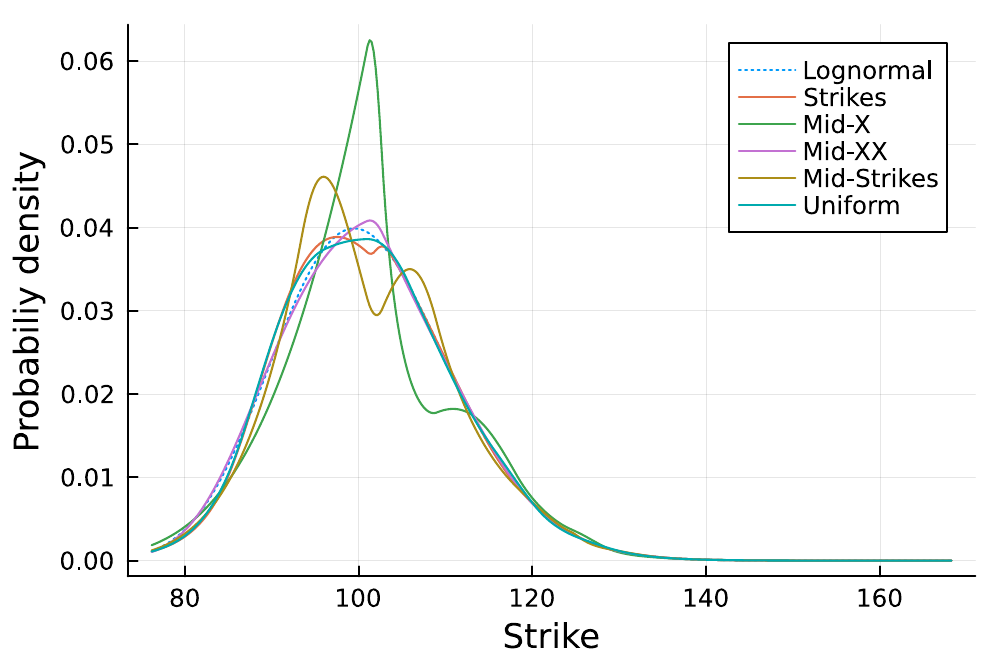} }\\
			\subfigure[\label{fig:lvgq_lognormal_density_set_C}Set C.]{
			\includegraphics[width=0.5\textwidth]{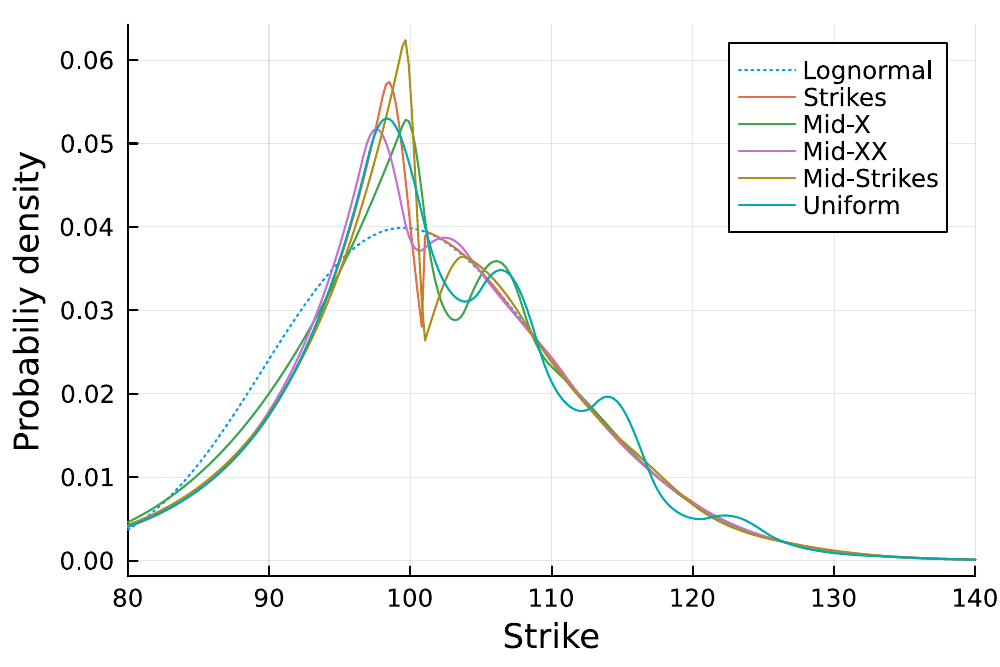}}
		\subfigure[\label{fig:lvgq_lognormal_density_set_D}Set D.]{
			\includegraphics[width=0.5\textwidth]{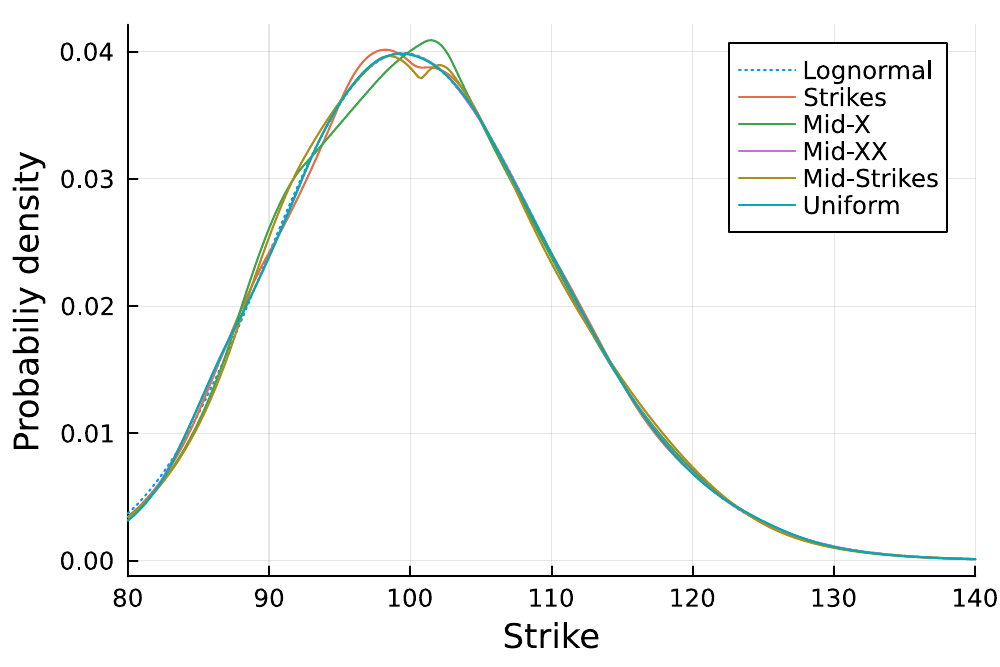} }
		\caption{Implied probability density of the calibrated quadratic LVG model using different sets of knots.}
\end{figure}

In order to assess the various knots candidates, we consider the same example as in the previous section, but using a few different random sets of 10 strikes in the interval [85,140] and a forward price $F=101$ (Table \ref{tbl:lognormal_strikes_sets}).
\begin{table}[H]
	\caption{Sets of market strikes with a Black-Scholes volatility of 20\% for a maturity $T=0.25$ and forward $F=101$. \label{tbl:lognormal_strikes_sets}}
	\centering{
	\begin{tabular}{lcccccccccc}\toprule
		Set & $K_1$ & $K_2$ & $K_3$ & $K_4$ & $K_5$ & $K_6$ & $K_7$ & $K_8$ & $K_9$ & $K_{10}$\\\midrule
		A & 88.77 & 92.85 & 93.38 & 99.37 & 107.99 & 120.29 & 122.03 & 123.9 & 134.71 & 135.43\\
		B & 85.02 & 101.92 & 103.55 & 114.45 & 121.85 & 123.69 & 125.07 & 125.58 & 131.63 & 133.86\\
		C & 98.07 & 100.93 & 101.06 &106.88 & 109.12 & 110.93 & 119.76 & 119.83 & 132.19 & 138.27\\	
		D &  85.00 & 		90.00 &		95.00 &		100.00 &		101.00 &		105.00 &		110.00 &		115.00 &		120.00 &		130.00\\					
		\bottomrule
	\end{tabular}}
\end{table}
\begin{table}[H]
	\caption{Root mean square error in implied volatilities \% for various choices of B-spline knots. The reference volatilities are flat 20\%.\label{tbl:lognormal_sets_rmse}}
	\centering{
		\begin{tabular}{lrrrrr}\toprule
			Set & Strikes & Mid-Strikes & Mid-X & Mid-XX & Uniform \\\midrule
			A & 9.4e-8 & 6.0e-3 & 5.2e-3 &4.1e-8 & 4.8e-9\\
			B & 9.9e-9 & 2.8e-3 &5.8e-1 & 2.9e-6 & 9.9e-3\\
			C & 1.0e-6 & 1.9e-3 & 1.0e-2 & 1.1e-8 & 1.4e-3\\
			D & 4.1e-4 &8.1e-2 &4.1e-2 & 2.6e-5 & 5.0e-7\\\bottomrule
	\end{tabular}}
\end{table}
The uniform discretization may\footnote{In practice, market strikes are not randomly distributed, but according to multiples of a minimum strike width, with more strikes near the money. The uniform discretization may still be relevant if some regularization is added to the objective of the minimizer.} result in strong oscillations due to overfitting in places where no market strike is quoted as in the set A (Figure \ref{fig:lvgq_lognormal_density_set_A}). 

The "Mid-Strikes" choice leads to a strong oscillation around the forward in set B (Figure \ref{fig:lvgq_lognormal_density_set_B}). The "Mid-X" produces a somewhat awkward shape on the set B. When the forward is very close to some of the knots as in set C, the "Strikes", "Mid-Strikes" choices lead to a density with a sharp gradient near the forward, a feature not desirable (Figure \ref{fig:lvgq_lognormal_density_set_C}). When the forward is part of the market strikes, a small wiggle is visible at the forward for "Strikes" and "Mid-Strikes" (Figure \ref{fig:lvgq_lognormal_density_set_D}).

Finally it is also interesting to look at the overall root mean square error in implied volatilities for the different choices (Table \ref{tbl:lognormal_sets_rmse}). The "Strikes" and "Mid-XX" choices consistently result in a near-perfect\footnote{The error in volatility is always below one basis point.} fit.

Overall, the "Mid-XX" knots lead to the most stable probability density along with an excellent fit. 

\subsubsection{Many quotes, few parameters}
In \citep{lefloc2020arbitrage}, regularization is employed to ensure a smooth implied probability density when fitting to many, eventually noisy, market option quotes. An interesting simpler alternative is to use few knots/few parameters instead of as many parameters as market quotes: by limiting the number of free-parameters, we may avoid overfitting issues, and at the same time we reduce the number of dimensions of the problem, thus increasing stability and performance.
 Where to place the knots then? Based on the previous observations, we may choose knots such that the market strikes are equidistributed. Concretely, we use \begin{equation*}
 	\tilde{\bm{K}}=\left\{K_1, K_{1+j},K_{1+2j},...,K_{n}\right\}\,,
 \end{equation*} where $j = n/m$ with $m \leq n$ and use the "Mid-XX" knots on top of $\tilde{\bm{K}}$. It may happen that many market strikes are quoted in a narrow range, in which case the set could be adjusted with a minimum strike width, although we did not need this tweak on the market examples presented below.

We consider previously published examples of real-world implied volatility smiles of equity indices, equities or foreign exchange rates which present interesting characteristics. The first example, presented in Table \ref{tbl:spx500-1w}, is the case of SPX500 expiring on March 24, 2017 as March 16, 2017 (one week - 1w) that the SVI parametrization of \citet{gatheral2006volatility} fails to fit properly.

The second example presented in Table \ref{tbl:spx500_feb5mar7} consists in options on SPX500 expiring on March 7, 2018 as of February 5, 2018 (one month - 1m) where the implied volatility smile exhibits a very high curvature, where the polynomial stochastic collocation leads to a somewhat unnatural spike in the probability density while a mixture of lognormals leads to a multimodal implied density \citep{lefloch2019model1}.

The third example uses very recent option quotes on TSLA of maturity March 21, 2025 as of February 21, 2025 (1m). The reference implied volatilities corresponding to the mid option prices are given in Table \ref{tbl:tsla_vols_1m}. The peculiarity of TSLA stock is that the volatility is relatively high. On this example, SVI does not fit the volatilities close to the money very well, while the xSSVI of \citet{corbetta2019robust} and SABR from \citet{hagan2002managing} (not shown for clarity) would be even worse.

The fourth example shows a W shaped implied volatility, typical before major earning events, for options on AAPL with expiry in 4 days as of October 28, 2013 (4d) from \citet{alexiou2021pricing}. Simple parametric models such as SVI or SABR or even polynomial stochastic collocation are unable to produce this kind of shape.

The fifth example (Table \ref{tbl:audnzd_quotes}) involves options on the AUD/NZD currency pair of maturity July 9, 2014 as of July 2, 2014 (1w) from \citet{wystup2018arbitrage}, where a typical calibration of SVI leads to a negative implied probability density.

There is an apparent focus on short maturities as those are more difficult to capture, with a variety of smile shapes, with some exhibiting multi-modality. As the implied volatility smile flattens for long maturities, those are much easier to fit to.

\begin{figure}[H]
		\subfigure[\label{fig:lvgq_spx500_1w_vol}Implied volatility.]{
			\includegraphics[width=0.5\textwidth]{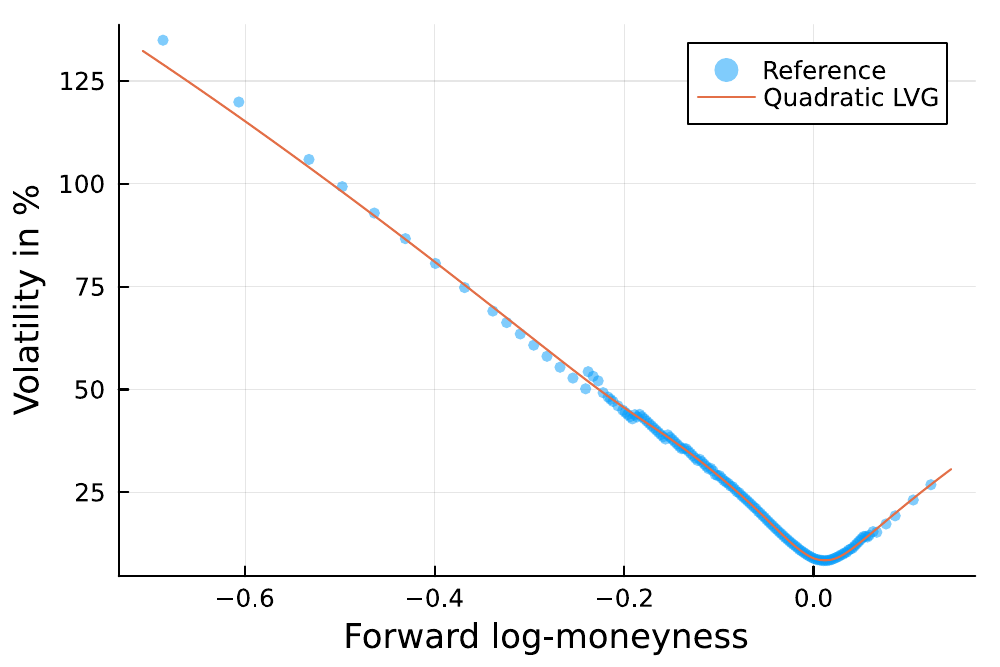}}
		\subfigure[\label{fig:lvgq_spx500_1w_dens}Implied probability density in log scale.]{
			\includegraphics[width=0.5\textwidth]{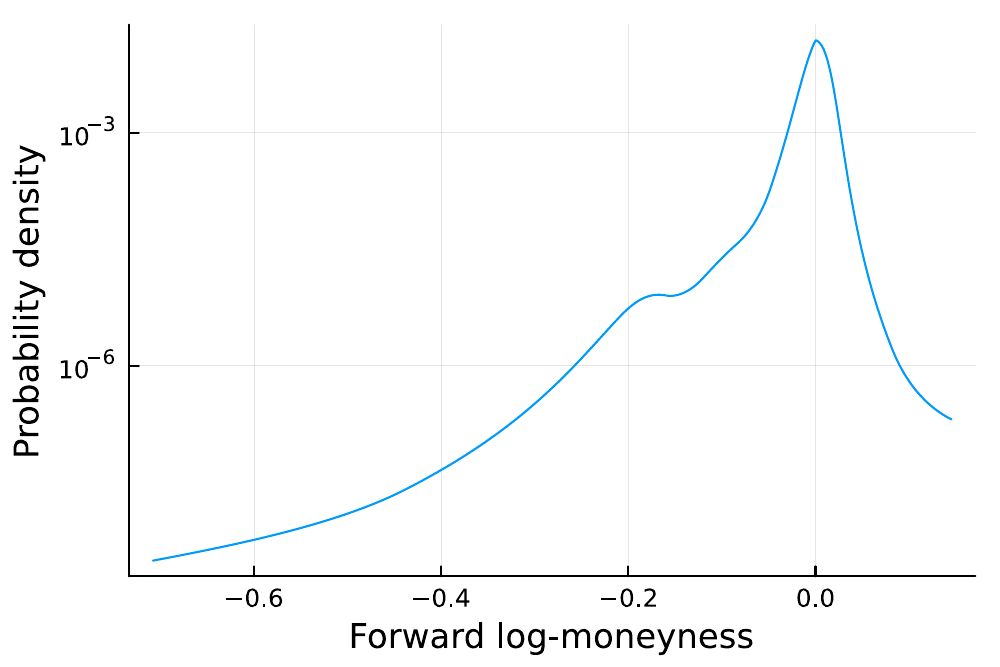} }\\
		\caption{Quadratic LVG model with 10 points calibrated to SPX500 options of maturity 1w.}
\end{figure}
\begin{figure}[H]
		\subfigure[\label{fig:lvgq_spx500_1m_vol}Implied volatility.]{
			\includegraphics[width=0.5\textwidth]{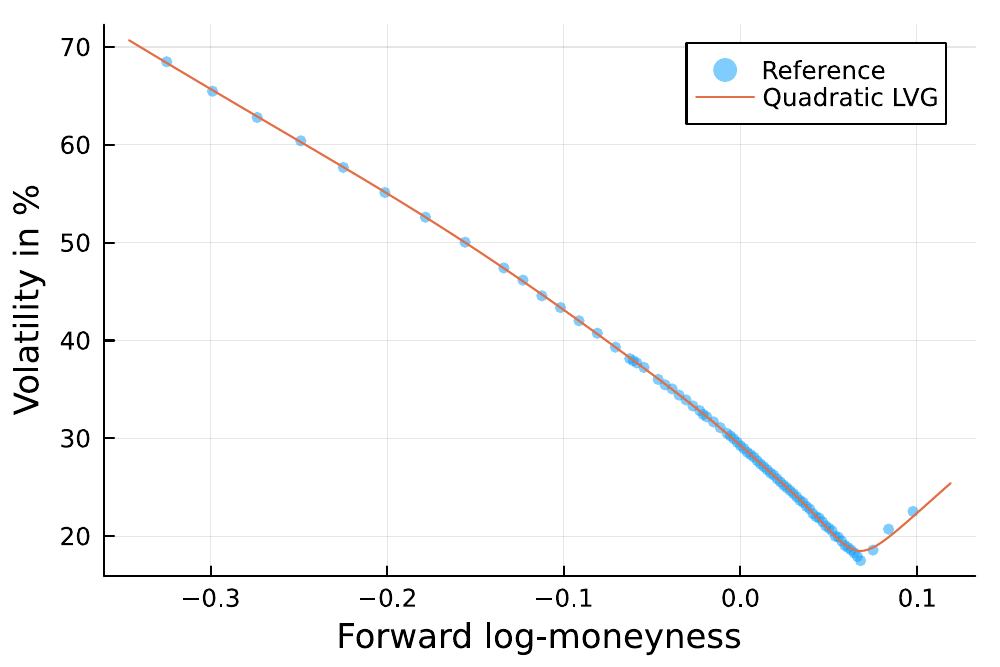}}
		\subfigure[\label{fig:lvgq_spx500_1m_dens}Implied probability density.]{
			\includegraphics[width=0.5\textwidth]{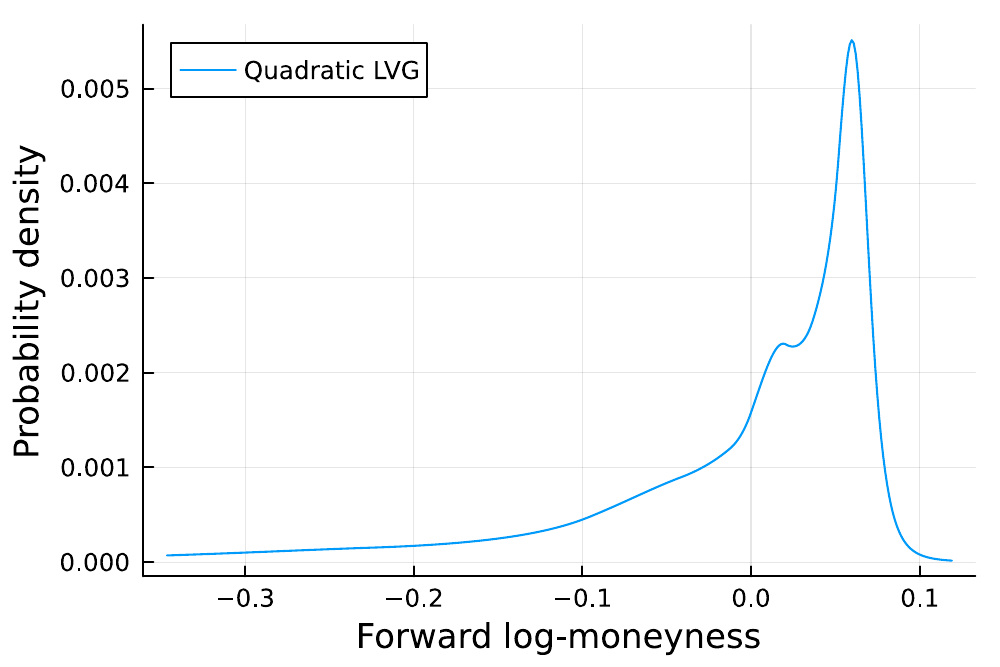} }\\
		\caption{Quadratic LVG model with 10 points calibrated to SPX500 options of maturity 1m.}
\end{figure}

\begin{figure}[H]
		\subfigure[\label{fig:lvgq_tsla_1m_vol}Implied volatility.]{
			\includegraphics[width=0.5\textwidth]{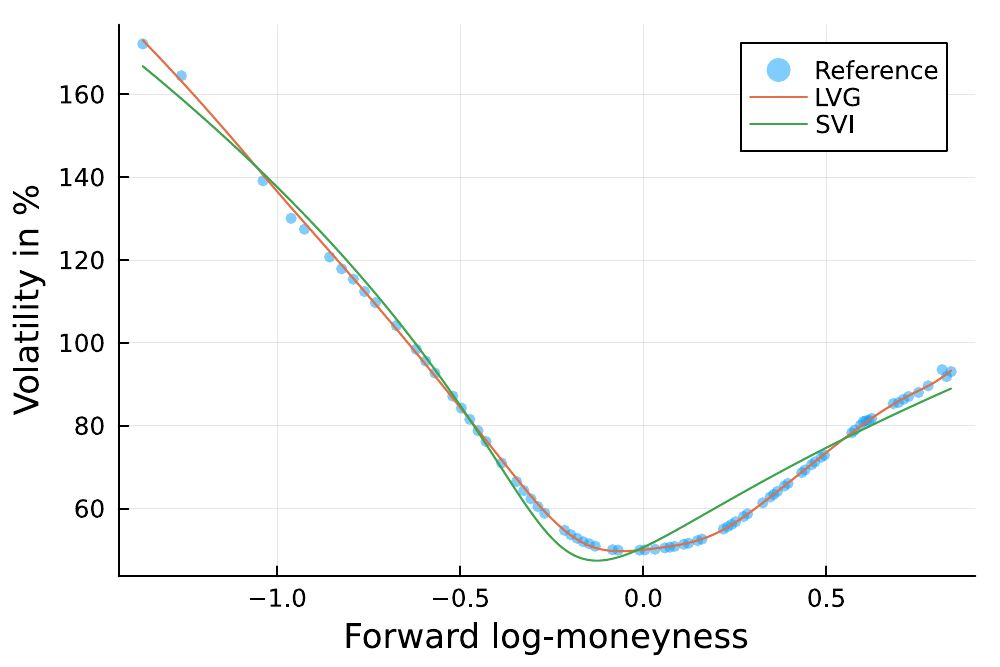}}
		\subfigure[\label{fig:lvgq_tsla_1m_dens}Implied probability density.]{
			\includegraphics[width=0.5\textwidth]{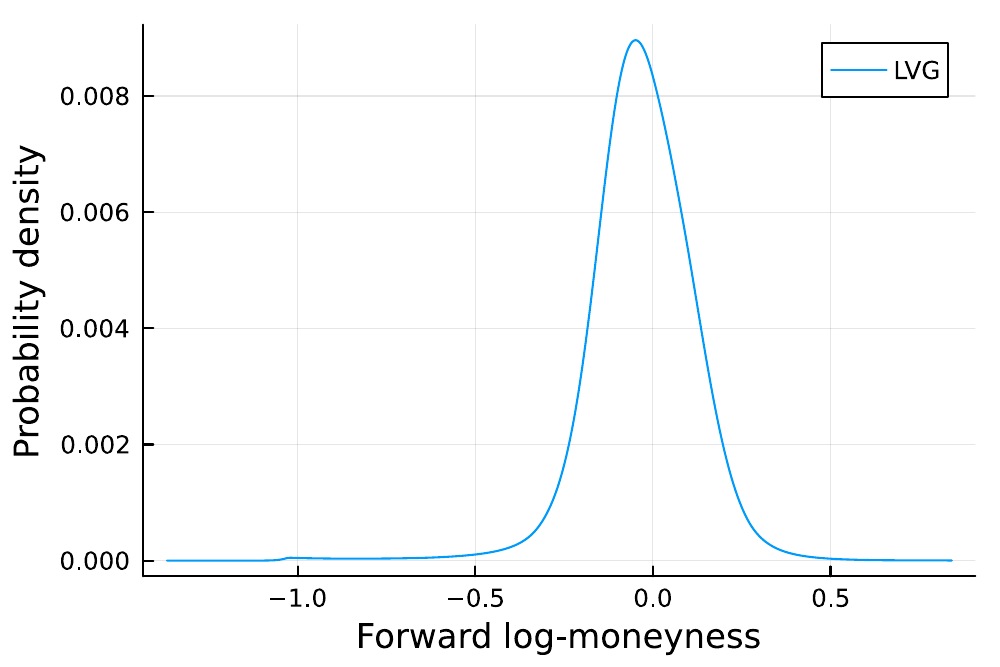} }\\
		\caption{Quadratic LVG model with 10 points calibrated to TSLA options of maturity 1m.}
\end{figure}
\begin{figure}[H]
		\subfigure[\label{fig:lvgq_aapl_4d_vol}Implied volatility.]{
			\includegraphics[width=0.5\textwidth]{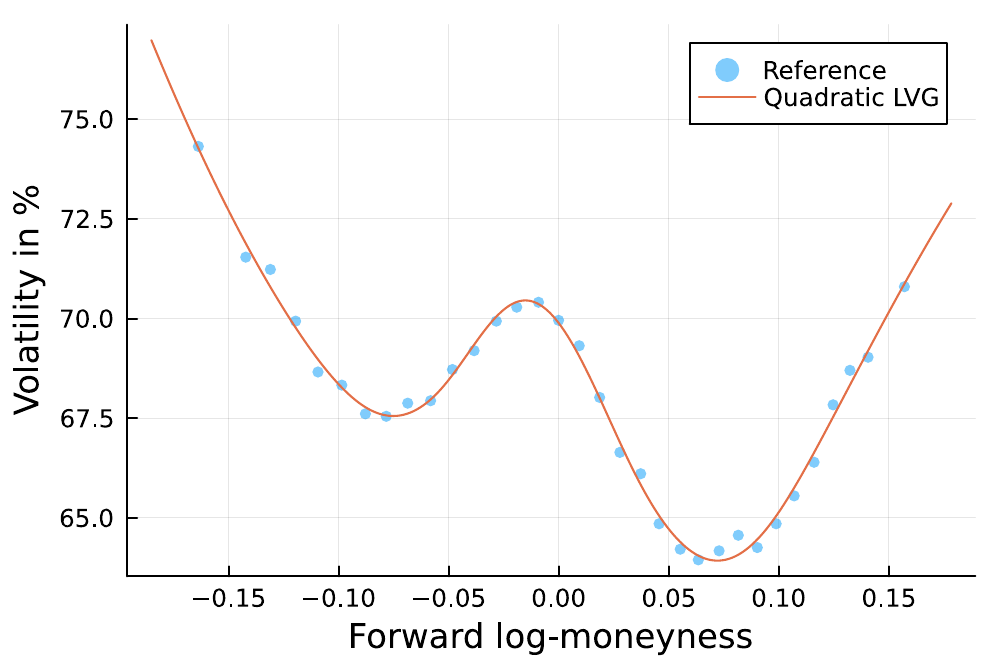}}
		\subfigure[\label{fig:lvgq_aapl_4d_dens}Implied probability density.]{
			\includegraphics[width=0.5\textwidth]{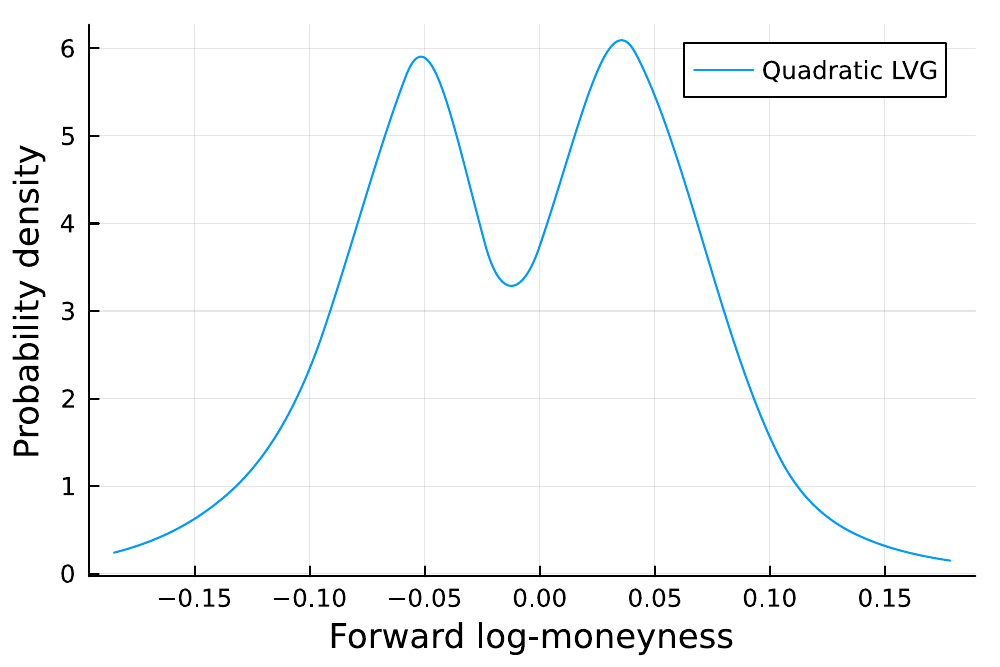} }\\
		\caption{Quadratic LVG model with 10 points calibrated to AAPL options of maturity 4d.}
\end{figure}

\begin{figure}[H]
		\subfigure[\label{fig:lvgq_audnzd_1w_vol}Implied volatility.]{
			\includegraphics[width=0.5\textwidth]{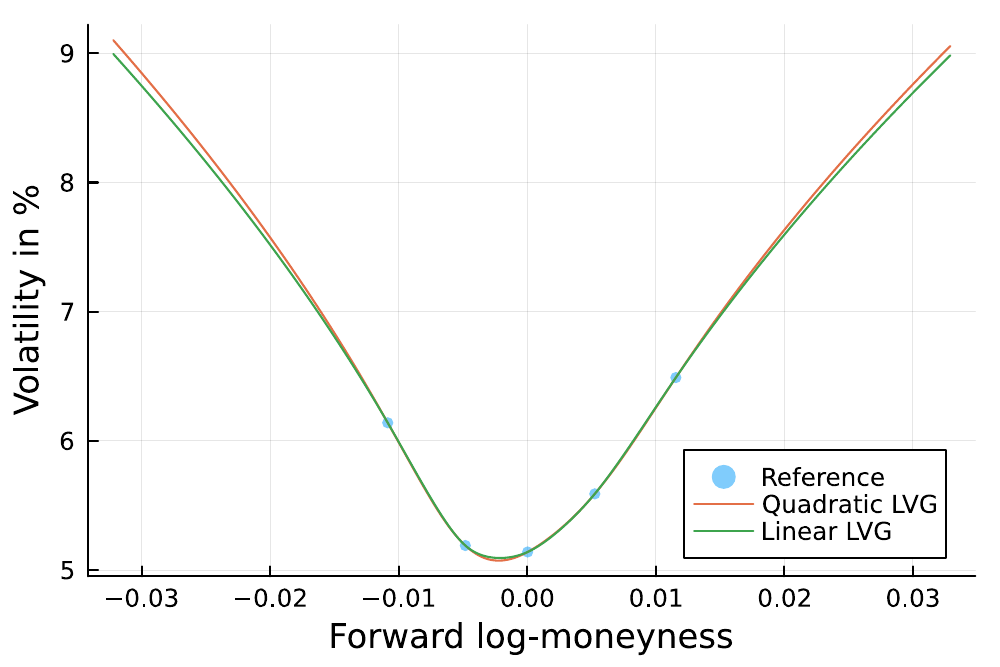}}
		\subfigure[\label{fig:lvgq_audnzd_1w_dens}Implied probability density.]{
			\includegraphics[width=0.5\textwidth]{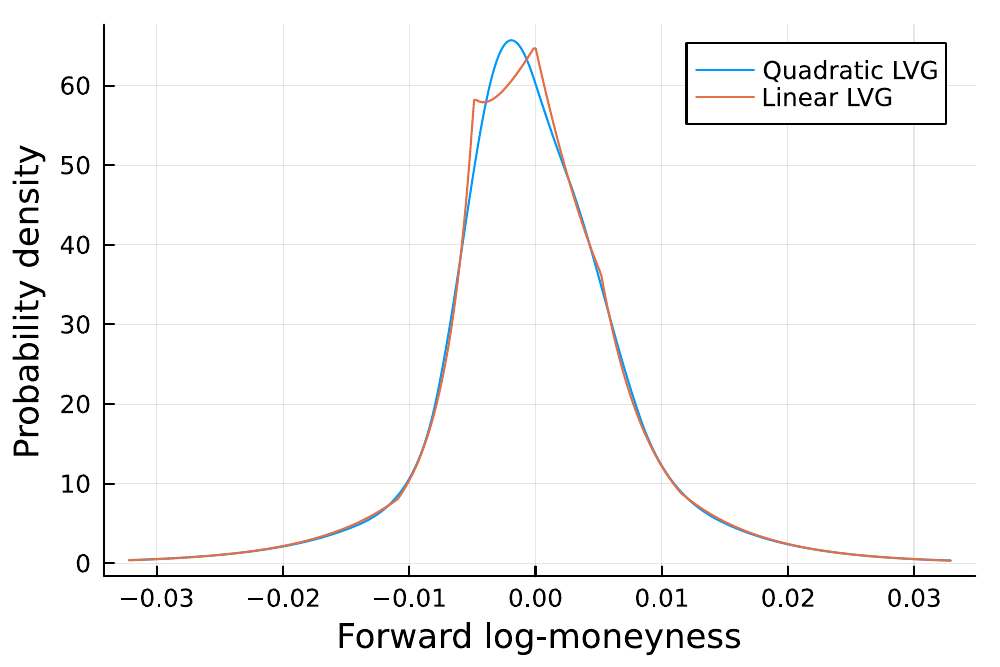} }\\
		\caption{Quadratic LVG model with 5 points calibrated to AUD/NZD options of maturity 1w.}
\end{figure}

 The AUD/NZD foreign exchange smile from \citet{wystup2018arbitrage} is useful to see how the local variance gamma model behaves on a minimalistic example: indeed, as is usual on the foreign exchange options market, it involves only five options quotes. In this case we use as many parameters as market quotes. Figure \ref{fig:lvgq_audnzd_1w_dens} shows that the density implied by the quadratic LVG model is smooth, but the one implied by the linear Bachelier LVG model exhibits some sharp unnatural gradients near the money.
Indeed, the linear LVG model leads only to a $\mathcal{C}^0$ probability density. Such gradients are then inevitable when the number of quotes is small.  On this example, the (unconstrained) SVI model is known to lead to  some negative probability density.

In all the examples considered so far, the fit in terms of implied volatilities is excellent, and the implied probability density is smooth, without spurious peaks, although the number of points considered ($n=10$) is somewhat arbitrary.

\subsubsection{Challenging examples of exact interpolation}
We consider the manufactured examples of \citet{jackel2014clamping} presented in Table \ref{tbl:jackel_clamping_1}. In the first example, a cubic spline interpolation of option prices is known to produce oscillations in the implied volatility, while a cubic spline on the volatilities introduces spurious arbitrages. In the second example, some of the quotes are at the limit of arbitrage.

On those examples, some care need to be taken in the choice of the boundaries $L$ and $U$: they must be far away enough. We pick $L = K_1 / 2$ and $U =  2 K_m$ where $K_1$ is the smallest quoted strike and $K_m$ the largest.

\begin{figure}[H]
		\subfigure[\label{fig:lvgq_jackel1_vol}Implied volatility.]{
			\includegraphics[width=0.5\textwidth]{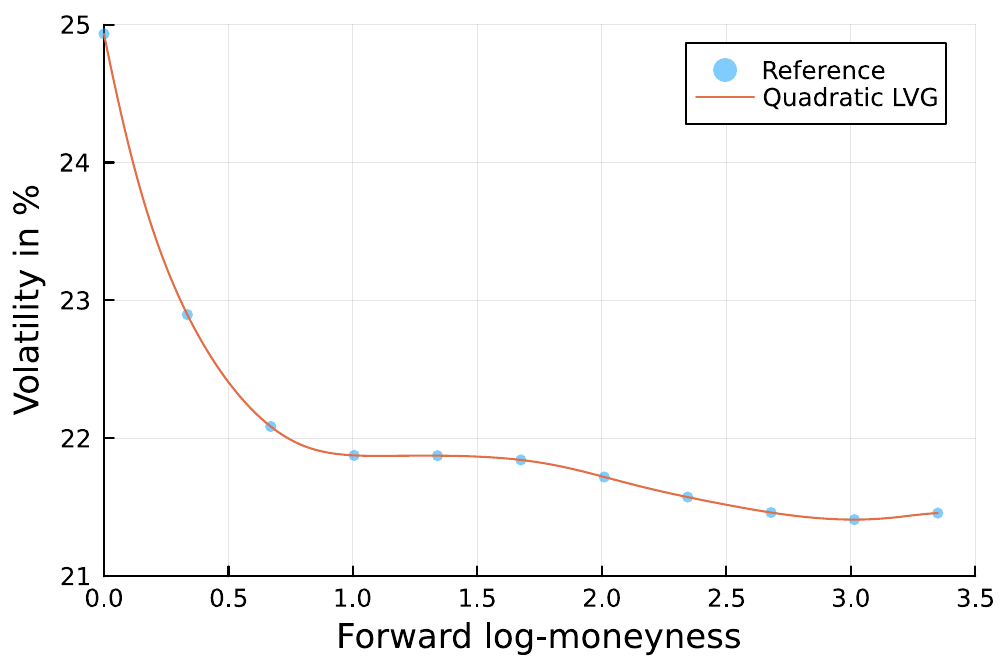}}
		\subfigure[\label{fig:lvgq_jackel1_dens}Implied probability density in logarithmic scale.]{
			\includegraphics[width=0.5\textwidth]{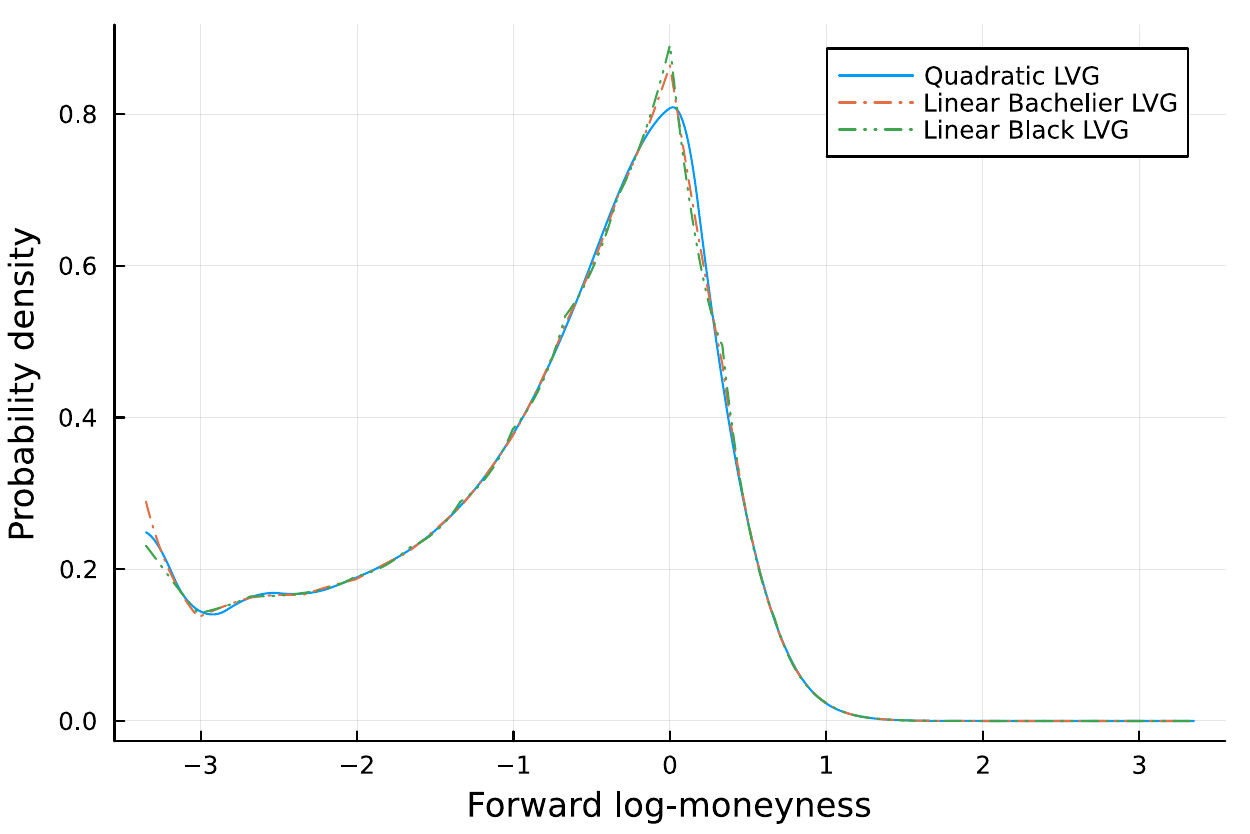} }\\
		\caption{LVG models calibrated to the example case I of \citet{jackel2014clamping}.}
\end{figure}

On the example case I, the fit is nearly exact for the linear Bachelier, Black and quadratic LVG models and there is no oscillation or wiggles in the implied volatility interpolation (Figure \ref{fig:lvgq_jackel1_vol}). The corresponding implied probability density is of course smoothest with the quadratic LVG model (Figure \ref{fig:lvgq_jackel1_dens}).

\begin{figure}[H]
		\subfigure[\label{fig:lvgq_jackel2_vol}Implied volatility.]{
			\includegraphics[width=0.5\textwidth]{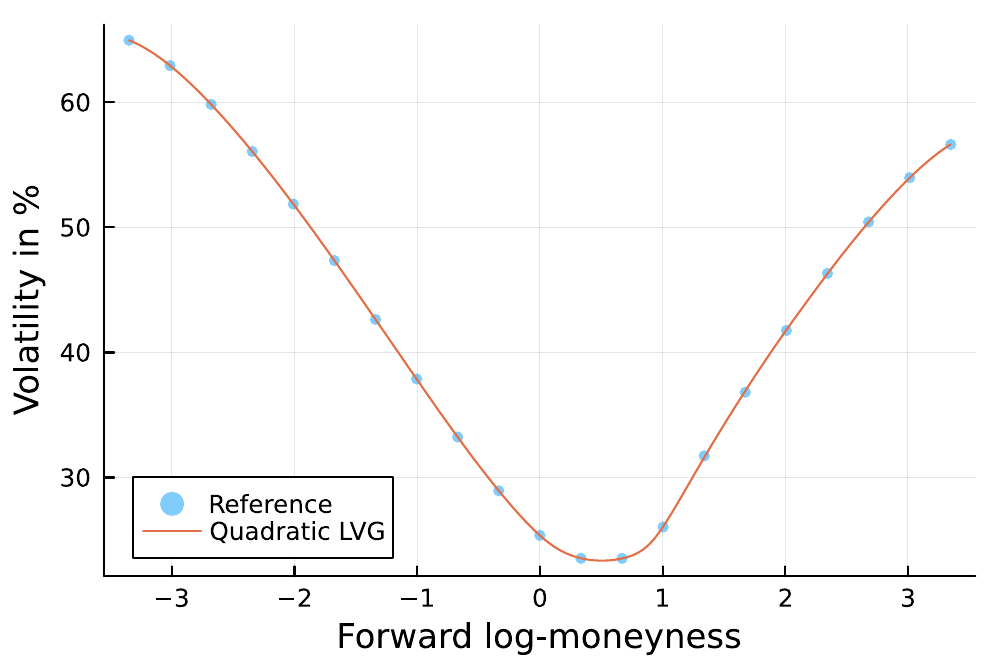}}
		\subfigure[\label{fig:lvgq_jackel2_dens}Implied probability density in logarithmic scale.]{
			\includegraphics[width=0.5\textwidth]{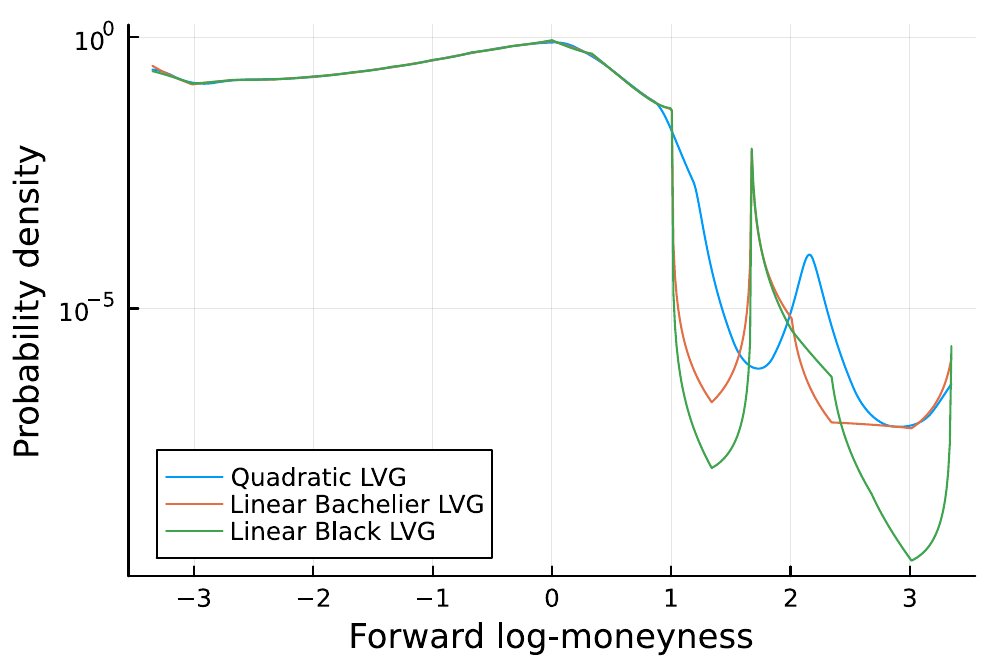} }\\
		\caption{LVG models calibrated to the example case II of \citet{jackel2014clamping}.}
\end{figure}

On the example case II, the quadratic LVG model does not allow for an exact fit. The root mean square error is around 4 basis points (Table \ref{tbl:jackel_lvgq_rmse}), the fit is qualitatively good (Figure \ref{fig:lvgq_jackel2_vol}). The near-arbitrages impose the probability density to go to almost zero, which conflicts with  the $\mathcal{C}^1$ continuity constraints of parameterized B-spline  density. The density stays however smooth, and looks more natural than the clear overfit of the linear LVG models (Figure \ref{fig:lvgq_jackel2_dens}). 
Our choice of range for the $\bm{\alpha}$ parameter does not allow the linear Bachelier model to fit as well as the linear Black model. Increasing the range would make the two implied probability densities even more similar.

\begin{table}[H]
	\caption{Root mean square error (RMSE) in implied volatilities of the  LVG models calibrated to the market data of Table \ref{tbl:jackel_clamping_1}.\label{tbl:jackel_lvgq_rmse}}
\centering{	\begin{tabular}{lrrr}\toprule
 Model & Case I & Case II\\	\midrule	
 Linear Bachelier & 5.00e-13 &  4.54e-6 \\
 Linear Black & 3.64e-12 & 8.04e-8\\
Quadratic & 2.25e-12 & 4.02e-4\\\bottomrule
\end{tabular}}
\end{table}

\subsubsection{Multiple maturities - interpolation in time}
In order to fit multiple option maturities and obtain a continuous representation of the implied volatility surface in both strikes and expiry time, the simplest approach consists in calibrating each maturity independently and interpolate in between maturities in total variance. While often used by practitioners due to its simplicity, this approach may however introduce calendar spread arbitrages (or equivalently, areas of negative forward variance). A possible work-around is to calibrate from the shortest maturity to the longest, paying attention to eventually shift the reference (market) quotes upwards if the previous maturity leads to a calendar spread with the current maturity to calibrate. This still does not guarantee the absence of calendar spreads in between market strikes, and even at the market strikes if the calibrated prices differ significantly from the market prices.

\begin{figure}[H]
		\centering{
			\includegraphics[width=0.8\textwidth]{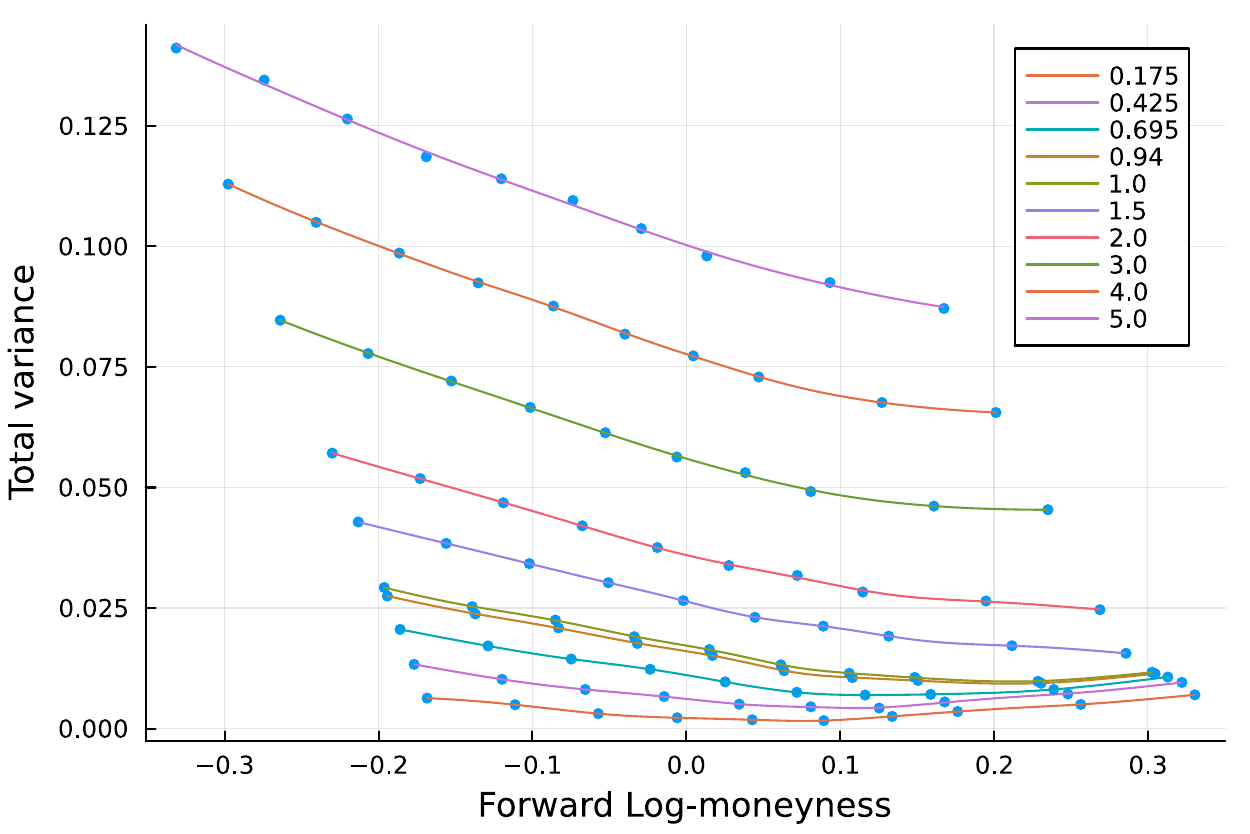}}
		\caption{Implied variance from the quadratic LVG model calibrated to the market data of \citet{kahale2004arbitrage}.\label{fig:kahale_lvg_totalvar}}	
	\end{figure}

\begin{figure}[H]
			\subfigure[Implied volatility.]{
				\includegraphics[width=0.5\textwidth]{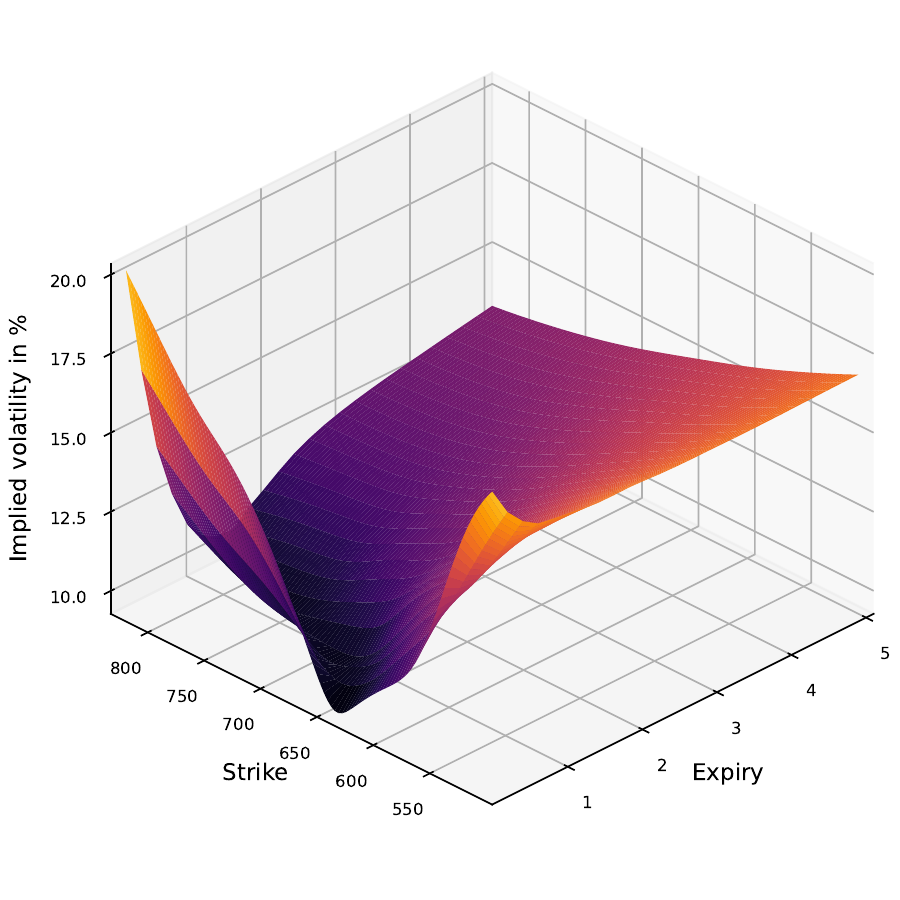}}
			\subfigure[Local volatility.]{
				\includegraphics[width=0.5\textwidth]{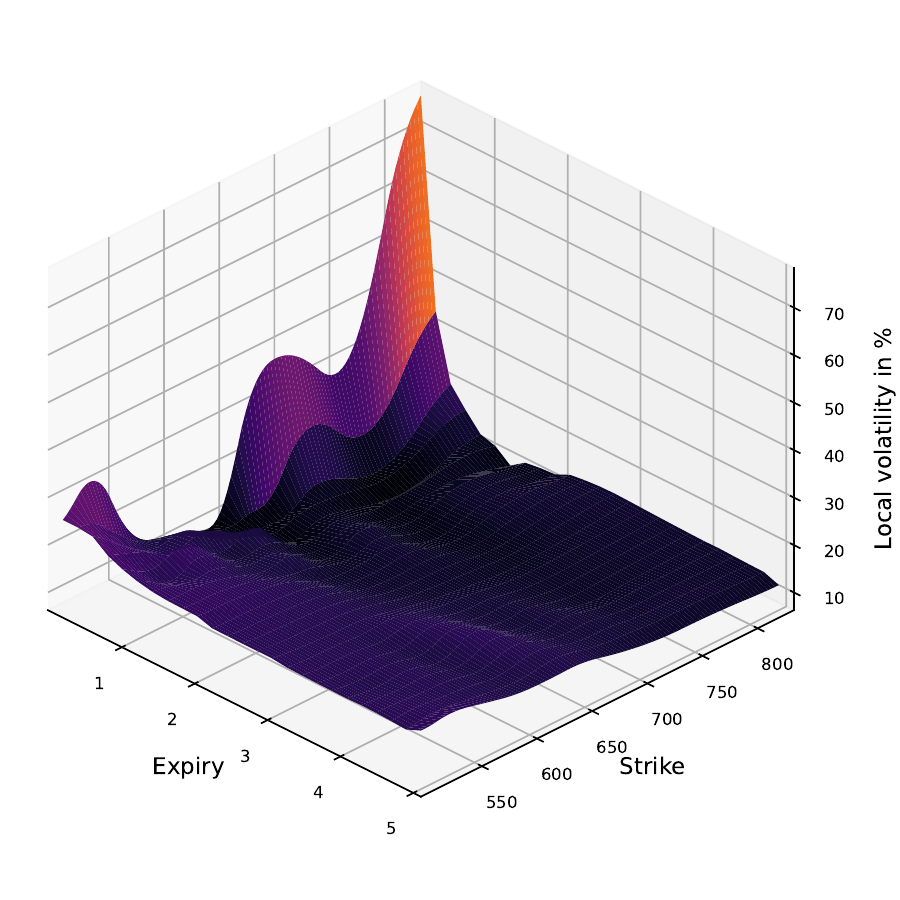} }\\
			\caption{Volatility surfaces from the quadratic LVG model calibrated to the market data of \citet{kahale2004arbitrage}.\label{fig:kahale_lvg_iv3d}}	
	\end{figure}

Another approach, suggested in \citep{falck2017local}, is to make sure the total local variance parameterization increases with the time to maturity. For the B-spline parameterization this requires to
\begin{itemize}
	\item use the same knots for all the market option maturities. To do so, we divide the prices and strikes by the forward at each maturity and use a forward $F(0,T_i)=1$ in the LVG calibration. This will provide a natural interpolation across constant forward moneyness. The original option price may be recovered by multiplying the model price by the forward at the relevant maturity. We also choose the "Mid-XX" knots of the first maturity, but we may use any maturity to base our knots on as they do not need to match any market strike.
	\item ensure that $\bm{\lambda}_{T_i} > \bm{\lambda}_{T_{i-1}}$ during the calibration instead of  $\bm{\lambda}_{T_i} > 0$ where $\bm{\lambda}_{T_i}$ are the B-spline parameters for the maturity $T_i$ and $T_i > T_{i-1}$. This is guarantees that the total local variance function $a(x)$ increases with the time to maturity for every $x \in [L,U]$, which is a sufficient condition to guarantee the absence of calendar spread arbitrages.	
\end{itemize}  
On the example market data of \citet{kahale2004arbitrage}, the above calibration method results in smooth implied variance curves. As expected from the absence of calendar spread arbitrages, the curves do not cross in Figure \ref{fig:kahale_lvg_totalvar}. The error in implied volatility is larger for the longer maturities, with up to 6 basis points for the last maturity. This may still be acceptable, and the increased smoothness is attractive. An additional global calibration step where all the expiries are calibrated together, based on the bootstrap calibration as initial guess, would allow to further improve the fit.
In between maturities, we may interpolate linearly the B-spline parameters in the square root of time:
\begin{equation*}
\bm{\lambda}_{t} = \bm{\lambda}_{T_{i-1}} + (\bm{\lambda}_{T_i}-\bm{\lambda}_{T_{i-1}})\frac{\sqrt{t-T_{i-1}}}{\sqrt{T_i-T_{i-1}}}\,,
\end{equation*} for $T_{i-1} < t \leq T_i$.
The corresponding "flat volatility" extrapolation would consist in using $\bm{\lambda}_{T_1} \sqrt{t / T_1}$ for $t < T_1$ and $\bm{\lambda}_{T_N} \sqrt{t / T_N}$ for $t > T_N$. Alternatively we may use a linear interpolation in total variance with flat implied volatility extrapolation.
The resulting Dupire local volatility surface is smooth (Figure \ref{fig:kahale_lvg_iv3d}). 

\section{Conclusion}
The quadratic local variance gamma model with a small number of knots lead to smooth implied probability densities on a variety of market option quotes, while providing an excellent fit in terms of implied volatilities, even on challenging examples of non convex implied volatilities or multi-modal probability densities.

It thus constitutes an interesting alternative to the non-parametric approaches of the linear or quadratic local variance gamma models or of the Andreasen-Huge one-step local volatility model, which all require a non-trivial choice of regularization constant to produce a smooth implied probability density.

In similar fashion as \citet{falck2017local}, it may be used as specific generic parameterization with a fixed, small number of parameters using only 3 or 5 points. In contrast to \citet{falck2017local}, where the linear parameterization leads to edges in the implied probability density and where the problem of the discontinuity at the forward is not dealt with, the probability density of the quadratic LVG model with a reduced number of points will be smooth.

Further research may explore the use of Fourier transforms to recover efficient pricing formulae for European options under the local variance gamma model with a more general local variance function.


\externalbibliography{yes}
\bibliography{arbfree_interpolation.bib}

\begin{thebibliography}{}

\bibitem[\protect\citeauthoryear{Alexiou, Goyal, Kostakis, and
  Rompolis}{Alexiou et~al.}{2021}]{alexiou2021pricing}
Alexiou, Lykourgos, Amit Goyal, Alexandros Kostakis, and Leonidas Rompolis.
  2021.
\newblock Pricing event risk: Evidence from concave implied volatility curves.
\newblock {\em Swiss Finance Institute Research Paper\/}~(21-48).

\bibitem[\protect\citeauthoryear{Andreasen and Huge}{Andreasen and
  Huge}{2011}]{andreasen2011volatility}
Andreasen, Jesper and Brian Huge. 2011.
\newblock Volatility interpolation.
\newblock {\em Risk\/}~{\em 24\/}(3), 76.

\bibitem[\protect\citeauthoryear{Breeden and Litzenberger}{Breeden and
  Litzenberger}{1978}]{breeden1978prices}
Breeden, Douglas~T and Robert~H Litzenberger. 1978.
\newblock Prices of state-contingent claims implicit in option prices.
\newblock {\em Journal of business\/}, 621--651.

\bibitem[\protect\citeauthoryear{Carr and Lee}{Carr and
  Lee}{2008}]{carr2008robust}
Carr, Peter and Roger Lee. 2008.
\newblock Robust replication of volatility derivatives.
\newblock In {\em Prmia award for best paper in derivatives, mfa 2008 annual
  meeting}.

\bibitem[\protect\citeauthoryear{Carr and Madan}{Carr and
  Madan}{2001}]{carr2001towards}
Carr, Peter and Dilip Madan. 2001.
\newblock Towards a theory of volatility trading.
\newblock {\em Option Pricing, Interest Rates and Risk Management, Handbooks in
  Mathematical Finance\/}, 458--476.

\bibitem[\protect\citeauthoryear{Carr and Nadtochiy}{Carr and
  Nadtochiy}{2017}]{carr2017local}
Carr, Peter and Sergey Nadtochiy. 2017.
\newblock Local variance gamma and explicit calibration to option prices.
\newblock {\em Mathematical Finance\/}~{\em 27\/}(1), 151--193.

\bibitem[\protect\citeauthoryear{Corbetta, Cohort, Laachir, and
  Martini}{Corbetta et~al.}{2019}]{corbetta2019robust}
Corbetta, Jacopo, Pierre Cohort, Ismail Laachir, and Claude Martini. 2019.
\newblock Robust calibration and arbitrage-free interpolation of ssvi slices.
\newblock {\em Decisions in Economics and Finance\/}~{\em 42\/}(2), 665--677.

\bibitem[\protect\citeauthoryear{De~Boor}{De~Boor}{1978}]{de1978practical}
De~Boor, Carl. 1978.
\newblock {\em A practical guide to splines}, Volume~27.
\newblock Springer-Verlag New York.

\bibitem[\protect\citeauthoryear{Dupire}{Dupire}{1994}]{dupire1994pricing}
Dupire, Bruno. 1994.
\newblock Pricing with a smile.
\newblock {\em Risk\/}~{\em 7\/}(1), 18--20.

\bibitem[\protect\citeauthoryear{Falck and Deryabin}{Falck and
  Deryabin}{2017}]{falck2017local}
Falck, Markus and Mikhail~Vladimirovich Deryabin. 2017.
\newblock Local variance gamma revisited.
\newblock {\em Available at SSRN 2659728\/}.

\bibitem[\protect\citeauthoryear{Gatheral}{Gatheral}{2006}]{gatheral2006volatility}
Gatheral, Jim. 2006.
\newblock {\em The volatility surface: a practitioner's guide}, Volume 357.
\newblock Wiley. com.

\bibitem[\protect\citeauthoryear{Hagan, Kumar, Lesniewski, and Woodward}{Hagan
  et~al.}{2002}]{hagan2002managing}
Hagan, Patrick~S, Deep Kumar, Andrew~S Lesniewski, and Diana~E Woodward. 2002.
\newblock Managing smile risk.
\newblock {\em Wilmott magazine\/}.

\bibitem[\protect\citeauthoryear{Healy}{Healy}{2019}]{healy2019applied}
Healy, Jherek. 2019.
\newblock {\em Applied Quantitative Finance for Equity Derivatives}.
\newblock Lulu.com.

\bibitem[\protect\citeauthoryear{J{\"a}ckel}{J{\"a}ckel}{2014}]{jackel2014clamping}
J{\"a}ckel, Peter. 2014.
\newblock Clamping down on arbitrage.
\newblock {\em Wilmott\/}~{\em 2014\/}(71), 54--69.

\bibitem[\protect\citeauthoryear{J{\"a}ckel}{J{\"a}ckel}{2015}]{jackel2013let}
J{\"a}ckel, Peter. 2015.
\newblock Let's be rational.

\bibitem[\protect\citeauthoryear{Kahal{\'e}}{Kahal{\'e}}{2004}]{kahale2004arbitrage}
Kahal{\'e}, Nabil. 2004.
\newblock An arbitrage-free interpolation of volatilities.
\newblock {\em Risk\/}~{\em 17\/}(5), 102--106.

\bibitem[\protect\citeauthoryear{Kanzow, Yamashita, and Fukushima}{Kanzow
  et~al.}{2004}]{kanzow2004levenberg}
Kanzow, Christian, Nobuo Yamashita, and Masao Fukushima. 2004.
\newblock Levenberg-marquardt methods for constrained nonlinear equations with
  strong local convergence properties.
\newblock {\em J. Computational and Applied Mathematics\/}~{\em 172}, 375--397.

\bibitem[\protect\citeauthoryear{Klare and Miller}{Klare and
  Miller}{2013}]{klare2013gn}
Klare, Kenneth and Guthrie Miller. 2013.
\newblock Gn--a simple and effective nonlinear least-squares algorithm for the
  open source literature.

\bibitem[\protect\citeauthoryear{{Le Floc'h}}{{Le
  Floc'h}}{2017}]{lefloch2017svi}
{Le Floc'h}, Fabien. 2017.
\newblock When {SVI} breaks down.
\newblock {\em
  \url{https://chasethedevil.github.io/post/when-svi-breaks-down}\/}.

\bibitem[\protect\citeauthoryear{Le~Floc’h}{Le~Floc’h}{2021}]{lefloc2020arbitrage}
Le~Floc’h, Fabien. 2021.
\newblock An arbitrage-free interpolation of class $\mathcal{C}^2$ for option
  prices.
\newblock {\em The Journal of Derivatives\/}~{\em 28\/}(4), 64--86.

\bibitem[\protect\citeauthoryear{Le~Floc’h and Oosterlee}{Le~Floc’h and
  Oosterlee}{2019a}]{lefloch2019model1}
Le~Floc’h, Fabien and Cornelis~W Oosterlee. 2019a.
\newblock Model-free stochastic collocation for an arbitrage-free implied
  volatility: Part i.
\newblock {\em Decisions in Economics and Finance\/}~{\em 42\/}(2), 679--714.

\bibitem[\protect\citeauthoryear{Le~Floc’h and Oosterlee}{Le~Floc’h and
  Oosterlee}{2019b}]{lefloch2019model2}
Le~Floc’h, Fabien and Cornelis~W Oosterlee. 2019b.
\newblock Model-free stochastic collocation for an arbitrage-free implied
  volatility, part ii.
\newblock {\em Risks\/}~{\em 7\/}(1), 30.

\bibitem[\protect\citeauthoryear{Wystup}{Wystup}{2018}]{wystup2018arbitrage}
Wystup, Uwe. 2018.
\newblock Arbitrage in the perfect volatility surface.
\newblock {\em Wilmott\/}~{\em 2018\/}(97), 16--17.

\end{thebibliography}
\appendixtitles{no}
\appendix
\section{Avoiding the use of complex numbers}\label{sec:avoid_complex_numbers}
We have $\chi_i =  \sqrt{\frac{a(x)}{\alpha_i}}$ and we know that $a(x) \geq 0$ since all the B-spline coefficients are positive, thus $\chi_i$ is either  real or imaginary if the sign  of $\alpha_i$ is positive (respectively strictly negative). Since $V(x)$ involves the ratio $\chi_i(x)/\chi_i(x_i)$, we may equivalently  calculate $\chi_i$ via $\chi_i=\sqrt{\frac{a(x)}{|\alpha_i|}}$.

From its definition, we know that $\omega_i$ is real if $\delta_i T + 8 \geq 0$ and imaginary otherwise.

For $z_i$ we need to consider the two cases where the roots $\tilde{x}_{i,1},\tilde{x}_{i,2}$ are both real or both complex. 

If they are real, we have $\frac{x-\tilde{x}_{i,1}}{x-\tilde{x}_{i,2}}= \frac{a(x)}{\alpha_i (x-\tilde{x}_{i,2})^2}$. We know that $a(x) \geq 0$ since the B-spline coefficients are all positive. If $\alpha_i > 0$ then $z_i$ is real, otherwise $z_i=\ln\left|\frac{x-\tilde{x}_{i,1}}{x-\tilde{x}_{i,2}}\right| + \iu \pi$.

If the roots are complex with non null imaginary parts, $\tilde{x}_{i,1},\tilde{x}_{i,2}$ are conjugates of each other. We may write $\tilde{x}_{i,1} = u+\iu v$ and $\tilde{x}_{i,2}
= u - \iu v$ with $(u,v) \in \mathbb{R}\times\mathbb{R}^{\star}$. We have then $\left|\frac{x-\tilde{x}_{i,1}}{x-\tilde{x}_{i,2}}\right| = \left|\frac{x-u-\iu v}{x-u+ \iu v} \right| = 1$ and thus $z_i$ is imaginary.

The term  $\cosh\left(\omega_i\left(z_i(x)-z_i(x_i)\right)\right)$ is thus always real. When $\omega_i\left(z_i(x)-z_i(x_i)\right)$ is imaginary, the term $\sinh\left(\omega_i\left(z_i(x)-z_i(x_i)\right)\right)$ is imaginary.

To evaluate the expression $V(x)$, there are three cases two consider:
\begin{itemize}
	\item $\delta_i \geq 0$ and $\delta_i T + 8 \geq 0$. All the variables are real.
	\item $\delta_i < 0$ and $\delta_i T + 8 \geq 0$. The logarithm in $z_i$ may be calculated using $\Im\left(z_i(x)\right)=\arcsin\left[\Im\left(\frac{x-\tilde{x}_{i,1}}{x-\tilde{x}_{i,2}}\right)\right]$. The imaginary part may be directly calculated (without complex numbers). Then use $\cos\left[\bar{\omega}_i \Im\left(z_i(x)\right)\right]$, $\sin\left[-\bar{\omega}_i \Im\left(z_i(x)\right)\right]$ instead of $\cosh$ and $\sinh$ terms. $\Theta_i^s$ will then be real instead of complex.
	\item $\delta_i < 0$ and $\delta_i T + 8 < 0$.  The logarithm in $z_i$ may be calculated using $\Im\left(z_i(x)\right)=\arcsin\left[\Im\left(\frac{x-\tilde{x}_{i,1}}{x-\tilde{x}_{i,2}}\right)\right]$. The imaginary part may be directly calculated (without complex numbers). Then use $\cosh\left[\bar{\omega}_i \Im\left(z_i(x)\right)\right]$, $\sinh\left[-\bar{\omega}_i \Im\left(z_i(x)\right)\right]$.
	
\end{itemize}

\section{Market data}
\begin{scriptsize}
	\begin{longtable}[H]{@{}rrrrrrrr@{}}
		\caption{Implied volatilities SPX500 European options expiring on March 24, 2017 (one week, $T=0.021918$) assuming proportional dividends, with $r=0.9\%$ and implied forward $F(T)=2385.103981$. $y$ represents the log-moneyness $\ln\frac{K}{F}$ taken from \citet{lefloch2017svi}, also in \citet{healy2019applied}.\label{tbl:spx500-1w}}
		\endfirsthead
		\endhead
		\toprule		
		{Strike} & {$y$} & {Volume} & {Vol} &{Call Bid} & {Call Ask} & {Put Bid} & {Put Ask} \\ \midrule
		1800 & -0.2815 & 0 & 0.58 & 0.00 & 1.24 & 0.00 & 0.61 \\ 
		1850 & -0.2541 & 3300 & 0.53 & 0.00 & 1.14 & 0.00 & 0.56 \\ 
		1875 & -0.2406 & 3402 & 0.50 & 0.00 & 1.09 & 0.00 & 0.53 \\ 
		1880 & -0.2380 & 1551 & 0.54 & 0.00 & 1.08 & 0.52 & 0.56 \\ 
		1890 & -0.2327 & 200 & 0.53 & 0.00 & 1.06 & 0.51 & 0.55 \\ 
		1900 & -0.2274 & 200 & 0.52 & 0.00 & 1.04 & 0.50 & 0.53 \\ 
		1910 & -0.2221 & 0 & 0.49 & 0.00 & 1.02 & 0.00 & 0.52 \\ 
		1980 & -0.1861 & 3 & 0.43 & 0.00 & 0.89 & 0.42 & 0.45 \\ 
		1990 & -0.1811 & 2 & 0.43 & 0.00 & 0.87 & 0.41 & 0.45 \\ 
		2000 & -0.1761 & 12 & 0.42 & 0.00 & 0.86 & 0.40 & 0.44 \\ 
		2010 & -0.1711 & 1 & 0.41 & 0.00 & 0.84 & 0.39 & 0.43 \\ 
		2030 & -0.1612 & 11 & 0.39 & 0.00 & 0.80 & 0.37 & 0.41 \\ 
		2035 & -0.1587 & 1 & 0.38 & 0.00 & 0.79 & 0.36 & 0.40 \\ 
		2050 & -0.1514 & 10 & 0.38 & 0.00 & 0.76 & 0.37 & 0.40 \\ 
		2055 & -0.1490 & 10 & 0.38 & 0.00 & 0.75 & 0.36 & 0.39 \\ 
		2085 & -0.1345 & 11 & 0.36 & 0.00 & 0.69 & 0.34 & 0.36 \\ 
		2090 & -0.1321 & 1 & 0.35 & 0.00 & 0.69 & 0.34 & 0.36 \\ 
		2100 & -0.1273 & 105 & 0.34 & 0.00 & 0.67 & 0.33 & 0.35 \\ 
		2105 & -0.1249 & 1 & 0.33 & 0.00 & 0.66 & 0.32 & 0.34 \\ 
		2120 & -0.1178 & 6 & 0.32 & 0.00 & 0.63 & 0.32 & 0.33 \\ 
		2125 & -0.1155 & 111 & 0.32 & 0.00 & 0.62 & 0.31 & 0.33 \\ 
		2130 & -0.1131 & 1 & 0.31 & 0.00 & 0.61 & 0.30 & 0.32 \\ 
		2135 & -0.1108 & 11 & 0.31 & 0.00 & 0.61 & 0.30 & 0.31 \\ 
		2140 & -0.1084 & 10 & 0.31 & 0.00 & 0.60 & 0.30 & 0.31 \\ 
		2150 & -0.1038 & 6610 & 0.29 & 0.00 & 0.53 & 0.29 & 0.30 \\ 
		2155 & -0.1015 & 1 & 0.29 & 0.00 & 0.52 & 0.28 & 0.30 \\ 
		2160 & -0.0991 & 8 & 0.29 & 0.00 & 0.51 & 0.28 & 0.30 \\ 
		2165 & -0.0968 & 2 & 0.28 & 0.00 & 0.50 & 0.28 & 0.29 \\ 
		2175 & -0.0922 & 3 & 0.27 & 0.00 & 0.48 & 0.27 & 0.28 \\ 
		2180 & -0.0899 & 66 & 0.27 & 0.00 & 0.47 & 0.27 & 0.28 \\ 
		2185 & -0.0876 & 15 & 0.27 & 0.00 & 0.46 & 0.26 & 0.27 \\ 
		2190 & -0.0853 & 26 & 0.26 & 0.00 & 0.45 & 0.26 & 0.27 \\ 
		2195 & -0.0831 & 13 & 0.26 & 0.00 & 0.44 & 0.25 & 0.26 \\ 
		2200 & -0.0808 & 113 & 0.25 & 0.00 & 0.43 & 0.25 & 0.25 \\ 
		2205 & -0.0785 & 13 & 0.25 & 0.00 & 0.42 & 0.25 & 0.25 \\ 
		2210 & -0.0762 & 14 & 0.24 & 0.00 & 0.41 & 0.24 & 0.25 \\ 
		2215 & -0.0740 & 102 & 0.24 & 0.00 & 0.41 & 0.23 & 0.24 \\ 
		2220 & -0.0717 & 2 & 0.23 & 0.00 & 0.40 & 0.23 & 0.24 \\ 
		2225 & -0.0695 & 43 & 0.23 & 0.00 & 0.39 & 0.22 & 0.23 \\ 
		2230 & -0.0672 & 93 & 0.22 & 0.00 & 0.38 & 0.22 & 0.23 \\ 
		2235 & -0.0650 & 2 & 0.22 & 0.00 & 0.37 & 0.21 & 0.22 \\ 
		2240 & -0.0628 & 134 & 0.21 & 0.00 & 0.36 & 0.21 & 0.22 \\ 
		2245 & -0.0605 & 11 & 0.21 & 0.00 & 0.35 & 0.21 & 0.21 \\ 
		2250 & -0.0583 & 104 & 0.20 & 0.00 & 0.34 & 0.20 & 0.21 \\ 
		2255 & -0.0561 & 205 & 0.20 & 0.00 & 0.34 & 0.20 & 0.20 \\ 
		2260 & -0.0539 & 184 & 0.19 & 0.00 & 0.33 & 0.19 & 0.20 \\ 
		2265 & -0.0517 & 89 & 0.19 & 0.00 & 0.32 & 0.19 & 0.19 \\ 
		2270 & -0.0495 & 74 & 0.18 & 0.00 & 0.28 & 0.18 & 0.19 \\ 
		2275 & -0.0473 & 377 & 0.18 & 0.00 & 0.27 & 0.18 & 0.18 \\ 
		2280 & -0.0451 & 207 & 0.17 & 0.00 & 0.26 & 0.17 & 0.17 \\ 
		2285 & -0.0429 & 478 & 0.17 & 0.00 & 0.25 & 0.17 & 0.17 \\ 
		2290 & -0.0407 & 295 & 0.16 & 0.00 & 0.24 & 0.16 & 0.17 \\ 
		2295 & -0.0385 & 280 & 0.16 & 0.00 & 0.18 & 0.16 & 0.16 \\ 
		2300 & -0.0363 & 1106 & 0.15 & 0.00 & 0.18 & 0.15 & 0.16 \\ 
		2305 & -0.0342 & 377 & 0.15 & 0.00 & 0.17 & 0.15 & 0.15 \\ 
		2310 & -0.0320 & 793 & 0.14 & 0.00 & 0.17 & 0.14 & 0.15 \\ 
		2315 & -0.0298 & 567 & 0.14 & 0.09 & 0.16 & 0.14 & 0.14 \\ 
		2320 & -0.0277 & 864 & 0.14 & 0.09 & 0.15 & 0.13 & 0.14 \\ 
		2325 & -0.0255 & 569 & 0.13 & 0.10 & 0.15 & 0.13 & 0.13 \\ 
		2330 & -0.0234 & 567 & 0.13 & 0.10 & 0.14 & 0.13 & 0.13 \\ 
		2335 & -0.0212 & 260 & 0.12 & 0.10 & 0.14 & 0.12 & 0.12 \\ 
		2340 & -0.0191 & 407 & 0.12 & 0.10 & 0.13 & 0.12 & 0.12 \\ 
		2345 & -0.0170 & 506 & 0.12 & 0.10 & 0.13 & 0.11 & 0.12 \\ 
		2350 & -0.0148 & 1850 & 0.11 & 0.10 & 0.12 & 0.11 & 0.11 \\ 
		2355 & -0.0127 & 528 & 0.11 & 0.10 & 0.11 & 0.11 & 0.11 \\ 
		2360 & -0.0106 & 1134 & 0.10 & 0.10 & 0.11 & 0.10 & 0.11 \\ 
		2365 & -0.0085 & 436 & 0.10 & 0.10 & 0.10 & 0.10 & 0.10 \\ 
		2370 & -0.0064 & 1063 & 0.10 & 0.09 & 0.10 & 0.10 & 0.10 \\ 
		2375 & -0.0042 & 1420 & 0.09 & 0.09 & 0.10 & 0.09 & 0.10 \\ 
		2380 & -0.0021 & 1248 & 0.09 & 0.09 & 0.09 & 0.09 & 0.09 \\ 
		2385 & 0.0000 & 229 & 0.09 & 0.09 & 0.09 & 0.09 & 0.09 \\ 
		2390 & 0.0021 & 2252 & 0.09 & 0.09 & 0.09 & 0.09 & 0.09 \\ 
		2395 & 0.0041 & 912 & 0.09 & 0.09 & 0.09 & 0.08 & 0.09 \\ 
		2400 & 0.0062 & 2650 & 0.09 & 0.08 & 0.09 & 0.08 & 0.09 \\ 
		2405 & 0.0083 & 2523 & 0.08 & 0.08 & 0.09 & 0.08 & 0.09 \\ 
		2410 & 0.0104 & 789 & 0.08 & 0.08 & 0.09 & 0.08 & 0.10 \\ 
		2415 & 0.0125 & 863 & 0.08 & 0.08 & 0.08 & 0.07 & 0.10 \\ 
		2420 & 0.0145 & 905 & 0.08 & 0.08 & 0.09 & 0.00 & 0.13 \\ 
		2425 & 0.0166 & 1178 & 0.09 & 0.08 & 0.09 & 0.00 & 0.13 \\ 
		2430 & 0.0187 & 815 & 0.09 & 0.08 & 0.09 & 0.00 & 0.14 \\ 
		2435 & 0.0207 & 267 & 0.09 & 0.09 & 0.09 & 0.00 & 0.15 \\ 
		2440 & 0.0228 & 402 & 0.09 & 0.09 & 0.09 & 0.00 & 0.16 \\ 
		2445 & 0.0248 & 50 & 0.09 & 0.09 & 0.09 & 0.00 & 0.17 \\ 
		2450 & 0.0268 & 2027 & 0.09 & 0.09 & 0.10 & 0.00 & 0.17 \\ 
		2455 & 0.0289 & 59 & 0.10 & 0.10 & 0.10 & 0.00 & 0.18 \\ 
		2460 & 0.0309 & 159 & 0.10 & 0.10 & 0.10 & 0.00 & 0.19 \\ 
		2465 & 0.0330 & 26 & 0.10 & 0.10 & 0.11 & 0.00 & 0.20 \\ 
		2470 & 0.0350 & 53 & 0.10 & 0.10 & 0.11 & 0.00 & 0.21 \\ 
		2475 & 0.0370 & 229 & 0.11 & 0.10 & 0.11 & 0.00 & 0.22 \\ 
		2495 & 0.0450 & 1 & 0.12 & 0.11 & 0.13 & 0.00 & 0.25 \\ 
		2550 & 0.0669 & 800 & 0.15 & 0.00 & 0.16 & 0.00 & 0.37 \\ \bottomrule
	\end{longtable}
\end{scriptsize}

\begin{table}[H]
	\caption{Implied volatility quotes for SPX500 options expiring on March 7, 2018, as of February, 5, 2018. In ACT/365, the option maturity is $T=0.082192$. The implied forward price is $F(0,T)=2629.80$. The interest rate is $r=0.97\%$.\label{tbl:spx500_feb5mar7}}
	\centering{
		\small{
			\begin{tabular}{rrrlrrr}\toprule		
				\multicolumn{1}{l}{Strike} & \multicolumn{1}{l}{Logmoneyness} & \multicolumn{1}{l}{Implied vol.} &  & \multicolumn{1}{l}{Strike} & \multicolumn{1}{l}{Logmoneyness} & \multicolumn{1}{l}{Implied vol.} \\ \cmidrule(lr){1-3}\cmidrule(lr){5-7}
				1900 & -0.325055 & 0.684883 &  & 2650 & 0.007651 & 0.280853 \\ 
				1950 & -0.299079 & 0.6548 &  & 2655 & 0.009536 & 0.277035 \\ 
				2000 & -0.273762 & 0.627972 &  & 2660 & 0.011417 & 0.273715 \\ 
				2050 & -0.249069 & 0.604067 &  & 2665 & 0.013295 & 0.270891 \\ 
				2100 & -0.224971 & 0.576923 &  & 2670 & 0.01517 & 0.267889 \\ 
				2150 & -0.201441 & 0.551253 &  & 2675 & 0.017041 & 0.264533 \\ 
				2200 & -0.178451 & 0.526025 &  & 2680 & 0.018908 & 0.262344 \\ 
				2250 & -0.155979 & 0.500435 &  & 2685 & 0.020772 & 0.258598 \\ 
				2300 & -0.134 & 0.474137 &  & 2690 & 0.022632 & 0.2555 \\ 
				2325 & -0.123189 & 0.461716 &  & 2695 & 0.024489 & 0.25219 \\ 
				2350 & -0.112493 & 0.445709 &  & 2700 & 0.026343 & 0.249534 \\ 
				2375 & -0.101911 & 0.433661 &  & 2705 & 0.028193 & 0.246659 \\ 
				2400 & -0.09144 & 0.42016 &  & 2710 & 0.03004 & 0.243553 \\ 
				2425 & -0.081077 & 0.407463 &  & 2715 & 0.031883 & 0.240202 \\ 
				2450 & -0.070821 & 0.393168 &  & 2720 & 0.033723 & 0.236588 \\ 
				2470 & -0.062691 & 0.381405 &  & 2725 & 0.03556 & 0.234574 \\ 
				2475 & -0.060668 & 0.3793 &  & 2730 & 0.037393 & 0.230407 \\ 
				2480 & -0.05865 & 0.377109 &  & 2735 & 0.039223 & 0.227866 \\ 
				2490 & -0.054626 & 0.372471 &  & 2740 & 0.041049 & 0.223049 \\ 
				2510 & -0.046626 & 0.360294 &  & 2745 & 0.042872 & 0.219888 \\ 
				2520 & -0.04265 & 0.354671 &  & 2750 & 0.044692 & 0.218498 \\ 
				2530 & -0.03869 & 0.350533 &  & 2755 & 0.046509 & 0.214702 \\ 
				2540 & -0.034745 & 0.34419 &  & 2760 & 0.048322 & 0.210506 \\ 
				2550 & -0.030815 & 0.339273 &  & 2765 & 0.050132 & 0.208175 \\ 
				2560 & -0.026902 & 0.333069 &  & 2770 & 0.051939 & 0.205508 \\ 
				2570 & -0.023003 & 0.328206 &  & 2775 & 0.053742 & 0.199967 \\ 
				2575 & -0.021059 & 0.324314 &  & 2780 & 0.055542 & 0.199007 \\ 
				2580 & -0.019119 & 0.322041 &  & 2785 & 0.057339 & 0.195062 \\ 
				2590 & -0.015251 & 0.3168 &  & 2790 & 0.059133 & 0.190547 \\ 
				2600 & -0.011397 & 0.310914 &  & 2795 & 0.060923 & 0.188427 \\ 
				2610 & -0.007559 & 0.305042 &  & 2800 & 0.062711 & 0.185893 \\ 
				2615 & -0.005645 & 0.302416 &  & 2805 & 0.064495 & 0.182878 \\ 
				2620 & -0.003734 & 0.299488 &  & 2810 & 0.066276 & 0.179292 \\ 
				2625 & -0.001828 & 0.29609 &  & 2815 & 0.068053 & 0.175001 \\ 
				2630 & 7.5E-05 & 0.292378 &  & 2835 & 0.075133 & 0.185751 \\ 
				2635 & 0.001974 & 0.289516 &  & 2860 & 0.083913 & 0.207173 \\ 
				2640 & 0.00387 & 0.28584 &  & 2900 & 0.097802 & 0.225248 \\ 
				2645 & 0.005762 & 0.283342 &  & \multicolumn{1}{l}{} & \multicolumn{1}{l}{} & \multicolumn{1}{l}{} \\ \bottomrule
	\end{tabular}}}
\end{table}

\begin{small}
	\begin{longtable}[H]{@{}ccc  ccc@{}}
		\caption{Implied volatilities of TSLA American options expiring on March 21, 2025 as of February, 21 2025 ($T= 0.076712$) 
			with implied spot $S=351.9764$ and forward $F(T)=353.4459$, a discount rate $r=4.38987\%$.\label{tbl:tsla_vols_1m}}
		\endfirsthead
		\endhead
		\toprule	
		\multicolumn{3}{c}{Call} & \multicolumn{3}{c}{Put}\\
		\cmidrule(r){1-3}\cmidrule(l){4-6}
	Strike & Mid Price & Vol in \%& Strike & Mid Price & Vol in \%\\
355 & 18.725 & 50.00 & 90 & 0.050 & 172.19 \\
365 & 14.600 & 50.17 &100 & 0.070 & 164.49 \\
375 & 11.275 & 50.49 &125 & 0.085 & 139.13 \\
380 & 9.875 & 50.68 &135 & 0.090 & 130.07 \\
385 & 8.625 & 50.86 &140 & 0.105 & 127.45 \\
395 & 6.575 & 51.38 &150 & 0.125 & 120.75 \\
400 & 5.725 & 51.63 &155 & 0.140 & 117.89 \\
410 & 4.350 & 52.26 &160 & 0.160 & 115.39 \\
415 & 3.800 & 52.63 &165 & 0.175 & 112.39 \\
440 & 2.020 & 55.04 &170 & 0.195 & 109.71 \\
445 & 1.795 & 55.58 &180 & 0.235 & 104.16 \\
450 & 1.600 & 56.15 &190 & 0.275 & 98.47 \\
455 & 1.440 & 56.80 &195 & 0.295 & 95.61 \\
465 & 1.170 & 58.07 &200 & 0.315 & 92.73 \\
470 & 1.060 & 58.72 &210 & 0.360 & 87.15 \\
490 & 0.735 & 61.38 &215 & 0.380 & 84.26 \\
500 & 0.625 & 62.78 &220 & 0.405 & 81.53 \\
505 & 0.575 & 63.42 &225 & 0.430 & 78.79 \\
510 & 0.535 & 64.14 &230 & 0.460 & 76.17 \\
520 & 0.460 & 65.46 &240 & 0.525 & 70.98 \\
525 & 0.425 & 66.05 &250 & 0.635 & 66.51 \\
545 & 0.325 & 68.68 &255 & 0.700 & 64.32 \\
550 & 0.305 & 69.32 &260 & 0.785 & 62.33 \\
560 & 0.270 & 70.60 &265 & 0.890 & 60.47 \\
565 & 0.255 & 71.25 &270 & 1.030 & 58.85 \\
575 & 0.225 & 72.40 &285 & 1.685 & 54.74 \\
580 & 0.210 & 72.90 &290 & 2.025 & 53.69 \\
625 & 0.135 & 78.29 &295 & 2.455 & 52.81 \\
630 & 0.130 & 78.92 &300 & 2.975 & 52.00 \\
640 & 0.120 & 80.11 &305 & 3.650 & 51.49 \\
645 & 0.120 & 81.02 &310 & 4.425 & 50.90 \\
650 & 0.110 & 81.20 &325 & 7.850 & 50.05 \\
655 & 0.100 & 81.30 &330 & 9.400 & 49.96 \\
660 & 0.095 & 81.76 & 350 & 17.750 & 49.96 \\
700 & 0.065 & 85.33 & & &\\
710 & 0.055 & 85.59 &&&\\
720 & 0.050 & 86.37 &&&\\
730 & 0.045 & 87.05 &&&\\
750 & 0.035 & 88.03 &&&\\
770 & 0.030 & 89.63 &&&\\
800 & 0.030 & 93.53 &&&\\
810 & 0.020 & 91.87 &&&\\
820 & 0.020 & 93.08 &&&\\
		\bottomrule
	\end{longtable}
\end{small}

\begin{table}[H]
	\caption{Quotes for options on AUD/NZD currency pair of maturity July 9th, 2014 as of July 2nd, 2014. $S = 1.0784, F(0,T) = 1.07845, B(0,T) = 0.999712587139$, 10-$\Delta$ RR = 0.35\%,
		25-$\Delta$ RR = 0.40\%, ATM vol = 5.14\%, 25-$\Delta$ BF =
		0.25\%, 10-$\Delta$ BF = 1.175\%.\label{tbl:audnzd_quotes}}
	\centering{
		\begin{tabular}{cccccc}	
			\toprule
			Maturity & 10$\Delta$-Put & 25$\Delta$-Put & ATM & 25$\Delta$-Call & 10$\Delta$-Call \\\midrule
			1w &6.14\% & 5.19\% & 5.14\% & 5.59\%  &  6.49\% \\\bottomrule
	\end{tabular}}
\end{table}

\begin{table}[H]
	\centering{
		\caption{Black-Scholes implied volatilities against moneyness $\frac{x}{X(0)}$ for an option of maturity $T=5.0722$, examples 1 and 2 of \cite{jackel2014clamping}.}
		\begin{small}
			\begin{tabular}{rrr}\toprule
				\multicolumn{1}{c}{Moneyness} & \multicolumn{1}{c}{Volatility (Case I)} &  \multicolumn{1}{c}{Volatility (Case II)} \\ \midrule
				0.035123777453185 & 0.642412798191439 & 0.649712512502887 \\ 
				0.049095433048156 & 0.621682849924325 & 0.629372247414191\\ 
				0.068624781300891 & 0.590577891369241 & 0.598339248024188 \\ 
				0.095922580089594 & 0.553137221952525 & 0.560748840467284 \\ 
				0.134078990076508 & 0.511398042127817 & 0.518685454812697 \\ 
				0.18741338653678 & 0.466699250819768 & 0.473512707134552 \\ 
				0.261963320525776 & 0.420225808661573 & 0.426434688827871 \\ 
				0.366167980681693 & 0.373296313420122 & 0.378806875802102 \\ 
				0.511823524787378 & 0.327557513727855& 0.332366264644264 \\ 
				0.715418426368358 & 0.285106482185545 & 0.289407658380454\\ 
				1 & 0.249328882881654 & 0.253751752243855\\ 
				1.39778339939642 & 0.228967051575314 & 0.235378088110653 \\ 
				1.95379843162821 & 0.220857187809035 & 0.235343538571543 \\ 
				2.73098701349666 & 0.218762825294675 & 0.260395028879884 \\ 
				3.81732831143284 & 0.218742183617652 & 0.31735041252779 \\ 
				5.33579814376678 & 0.218432406892364 & 0.368205175099723 \\ 
				7.45829006788743 & 0.217198426268117 & 0.417582432865276 \\ 
				10.4250740447762 & 0.21573928902421 & 0.46323707706565 \\ 
				14.5719954372667 & 0.214619929462215 & 0.504386489988866 \\ 
				20.3684933182917 & 0.2141074555437 & 0.539752566560924 \\ 
				28.4707418310251 & 0.21457985392644  & 0.566370957381163\\ \bottomrule
			\end{tabular}
			\label{tbl:jackel_clamping_1}
	\end{small}}
\end{table}

\end{document}